\pdfoutput=1
\documentclass[%
 reprint,
 amsmath,amssymb,
 aps,
]{revtex4-1}

\usepackage{chemarr}
\usepackage{epsf,mathtools}
\usepackage{graphicx}
\usepackage{dcolumn}
\usepackage{bm}

\newcommand{\avg}[1]{\left\langle{#1}\right\rangle}

\begin{document}


\title{MicroRNAs as a selective channel of communication between\\ competing RNAs: a steady-state theory}

\author{
Matteo Figliuzzi$^{1}$, 
Enzo Marinari$^{1,\star}$,
Andrea De Martino$^{1,2,\star}$ 
}
\affiliation{%
$^1$ Dipartimento di Fisica, Sapienza Universit\`a di Roma, p.le A. Moro 2, 00185 Roma (Italy)\\
$^2$ IPCF-CNR, UoS Roma-Sapienza, Roma (Italy)\\
$^\star$These authors contributed equally to this work
}%


\begin{abstract}
It has recently been suggested that the competition for a finite pool of microRNAs (miRNA) gives rise to effective interactions among their common targets (competing endogenous RNAs or ceRNAs) that could prove to be crucial for post-transcriptional regulation (PTR). We have studied a minimal model of PTR where the emergence and the nature of such interactions can be characterized in detail at steady state. Sensitivity analysis shows that binding free energies and repression mechanisms are the key ingredients for the cross-talk between ceRNAs to arise. Interactions emerge in specific ranges of repression values, can  be symmetrical (one ceRNA influences another and vice-versa) or asymmetrical (one ceRNA influences another but not the reverse) and may be highly selective, while possibly limited by noise. In addition, we show that non-trivial correlations among ceRNAs can emerge in experimental readouts due to transcriptional fluctuations even in absence of miRNA-mediated cross-talk.
\end{abstract}

\pacs{Valid PACS appear here}
\maketitle


\section*{Introduction}

MicroRNAs (miRNAs) are 21-23 nucleotides (nt) long, endogeous, non-coding RNA molecules, that perform post-transcriptional regulation by specifically binding  target messenger RNAs (mRNAs), typically leading to a reduction in the levels of the corresponding proteins \cite{bartel,bartel2,mechanism}. They are transcribed from independent miRNA genes or from introns of protein-coding transcripts. After being processed into maturity, a miRNA is loaded onto a specialized class of proteins to form the RNA-induced Silencing Complex (RISC), which specifically binds miRNA response elements (MREs) located in target mRNAs (usually in their 3' UnTraslated Region or 3'-UTR) through a base-pairing recognition mechanism which requires at least 6-nt complementarity. The whole process, known as RNA interference (RNAi), results in gene silencing through translation inhibition and mRNA destabilization \cite{stoichiometric,mechanism}.

Each mRNA can typically interact with several miRNAs, and each miRNA can target many different mRNAs. Within the complex network of potential interactions that ensues, miRNAs have long been thought to function mainly as fine-tuners for regulation by weakly dampening the protein output \cite{bartel,flynt}. This view is supported by the fact that the statistically over-represented motifs (feed-forward or feed-back loops) that have been identified in the known miRNA--mRNA interaction network are indeed capable of buffering the noise level in the output layer (proteins)  \cite{tsang,re,caselle,wang}.  More recently the attention has been directed to system-level effects. In particular, it has been realized that miRNA-based regulation is strongly affected by global properties like the total concentration of available targets (a feature known as {\it dilution effect} \cite{marks}). 
The combination of the repressive effects of miRNAs on their targets and of the weakening of such repression due to dilution effects lead to effective, positive interactions between joint targets of a given miRNA {(\it cross-talk interactions)}.
In addition, it is now known that pseudo-genes and other long non-coding RNAs (lncRNAs) also possess MREs and can be bound by the RISC. This implies that, besides mRNAs, non-coding RNAs sharing identical MREs compete for common miRNAs \cite{pseudogene,cesana}. 
Thus, miRNAs appear to mediate the cross-talk between a broad class of competing endogenous RNAs (ceRNAs) which includes both mRNAs and lncRNAs commonly targeted by miRNAs, leading to a large-scale network of indirect interactions across the transcriptome \cite{salmena,sumazin}. Recent studies have shown that such interactions play a central role in many biological contexts, from muscle differentiation \cite{cesana} to cancer \cite{tay,karreth}.

Despite the body of experimental evidence, a clear quantitative understanding of miRNA-mediated regulation is still lacking. To address this issue, we formulate a minimal model of post-transcriptional regulation and analyze its steady state, aiming at quantifying the intensity of the interactions arising from competition through an analysis of the sensitivity to changes in the ceRNA transcription rates. We show that binding free energies and repression mechanisms are the key ingredients for the cross-talk between ceRNAs to arise. The emergent interactions can be symmetrical (one ceRNA influences another and vice-versa) or asymmetrical (one ceRNA influences another but not the reverse) and may be highly selective, although possibly hampered by noise.

We furthermore argue that the identification of cross-talk from gene expression data can be hindered by the fact that statistically significant correlations among ceRNAs can emerge in the experimental readouts simply due to transcriptional fluctuations.

\section*{Results}

\subsection*{One miRNA species, $N$ ceRNA species}

We start by considering a highly simplified system formed by $N$ different species of ceRNA molecules, labeled $m_i$ ($i=1,\ldots ,N$), targeted by a single miRNA species labeled $\mu$.  Each $m_i$ can reversibly bind $\mu$ in complexes labeled $c_i$.
 Conforming to the experimental evidence \cite{baek,kinetics}, we assume that translational repression is fast and precedes mRNA destabilization, implying that complexes cannot be translated and repression of translation simply occurs by sequestration of free ceRNAs. Taking a constant translation rate, the levels of unbound ceRNA can be used as a direct proxy for protein concentrations at steady state. The allowed processes with their respective rates are then as follows (see Fig. 1): 
\begin{figure}
\begin{center}
\includegraphics[width=8cm]{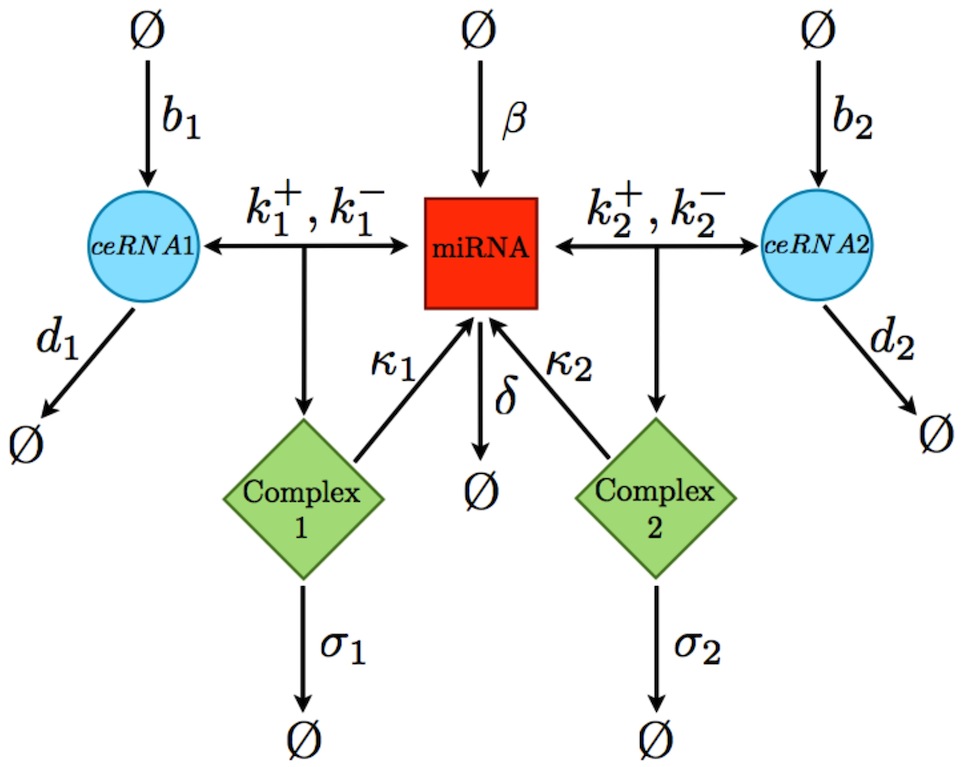}
\caption{Schematic representation of the considered processes for a system  $N=2$ ceRNA species and one miRNA.}
\end{center}
\end{figure}
\begin{gather*}
\emptyset \xrightleftharpoons[d_i]{b_i} m_i ~~~~~~~~~~~~
\emptyset \xrightleftharpoons[\delta]{\beta} \mu ~~~~~~~~~~~~
\mu+m_i \xrightleftharpoons[k_i^-]{k_i^+} c_i\\
c_i \xrightharpoonup{\sigma_i} \emptyset ~~~~~~~~~~~~
c_i \xrightharpoonup{\kappa_i} \mu 
\end{gather*}

Note that complexes are assumed to be degraded either through a catalytic channel (which gives back the miRNA to the cytosol) or through a stoichiometric channel (where both molecules are degraded with the complex). This choice serves the purpose of keeping the model as general as possible. 
The quantity $w_i=\sigma_i/(\sigma_i+\kappa_i)$  measures the degree of 'stoichiometricity' of complex decay: it ranges between $0$ (in case of fully catalytic degradation, $\sigma_i=0,\kappa_i>0$) and $1$ (in case of fully stoichiometric degradation, $\sigma_i>0,\kappa_i=0$). It will play an important role in our theory and we will refer to as {\it 'stoichiometricity ratio'.}
The exact mechanism of target repression is still a matter of debate, and in the past years several mechanisms have been reported \cite{bartel2,stoichiometric,mechanism}. Generically, miRNAs incorporated into the RISC do not seem to decay with their target, thus becoming again available for a new round of target RNA silencing. Nevertheless complexes may enter in specific cellular structure-like P-bodies thus resulting in an effective stoichiometric sequestration of both the miRNA and the target.

 Clearly, this setup represents a coarse-graining of the real biological processes, which typically requires multiple catalyzed elementary steps (e.g. in the formation of the RISC). However, such details may be disregarded if, in presence of many different targets, the only rate-limiting factor is the miRNA concentration. This will indeed be our main assumption (together with the fact that $m_i$'s can only interact through $\mu$). 

Denoting the concentration of species $x$ by $[x]$, we can write the mass-action kinetic rate equations for the above system as
\begin{gather}
\frac{d}{dt}[ m_i]=-d_i[ m_i] + b_i-k_i^+[ \mu][ m_i]+k_i^-[ c_i]\nonumber\\
\frac{d}{dt}[ \mu]=-\delta[ \mu] + \beta-\sum_i k_i^+[\mu][ m_i]+\sum_i(k_i^-+\kappa_i)[ c_i]\\
\frac{d}{dt}[ c_i]=-(\sigma_i+k_i^-+\kappa_i)[c_i] + k_i^+[ \mu][ m_i]\nonumber~~.
\end{gather}
In turn, they lead to the steady state equations
\begin{gather}\label{onemirna}
[m_i]=\frac{b_i+k_i^-[c_i]}{d_i+k_i^+[\mu]}\equiv m_i^\star F_i([\mu])\nonumber \\
[\mu]=\frac{\beta+\sum_i(k_i^-+\kappa_i)[c_i]}{\delta+\sum_i k_i^+[m_i]}\label{sseq}\\
[c_i]=\frac{ k_i^+[\mu][ m_i]}{\sigma_i+k_i^-+\kappa_i}\equiv c_i^\star \frac{[\mu]}{\mu_{0,i}} F_i([\mu])\nonumber~~.
\end{gather}
where we defined $m^\star_i=b_i/d_i$, $c^\star_i=b_i/(\sigma_i+\kappa_i)$ and
\begin{equation}\label{tre}
F_i([\mu])=\frac{\mu_{0,i}}{[\mu]+\mu_{0,i}}~~~~~,~~~~~
\mu_{0,i}=\frac{d_i}{k_i^+}(1+\phi_i)
\end{equation}
with $\phi_i=k_i^-/(\sigma_i+\kappa_i)$. Note that $m_i^\star$ and $c_i^\star$ represent the maximum  concentrations of free ceRNAs and complexes,  achievable in absence of miRNAs and in the limit of infinite miRNA concentration respectively.  A simple calculation shows that the binding free energy of the complex (in units of $k_B T$) is given by
\begin{equation}
\Delta G_i=-\log\frac{k_i^+ [m_i] [\mu]}{k_i^- [c_i]}=-\log\frac{1+\phi_i}{\phi_i}~~.
\end{equation}
This clarifies the physical meaning of $\phi_i$: if $\phi_i\gg 1$  then complex formation is close to equilibrium; if $\phi_i\ll 1$ instead, the process is unbalanced towards association. 

On the other hand, one sees that the quantity $\mu_{0,i}$ defined in (\ref{tre}) (which only depends on the kinetic parameters of ceRNA $m_i$) gives the miRNA level for which the concentrations of free ceRNAs and of complexes equal half of their theoretical maxima.  Therefore, in practice, we have the following situation:
\begin{enumerate}
\item[a.] If $[\mu]\ll\mu_{0,i}$ (or $[\mu]/\mu_{0,i}=\mathcal{O}(\epsilon)$), then $[m_i]\simeq m_i^\star$ and $[c_i]\ll c_i^\star$: here the levels of free ceRNAs are largest, while complexes are roughly absent; spontaneous degradation is the dominant channel of ceRNA decay.
\item[b.] If $[\mu]\simeq \mu_{0,i}$ (or $\left| 1-[\mu]/\mu_{0,i}\right| =\mathcal{O}(\epsilon)$), then $[m_i]\simeq m_i^\star/2$ and $c_i\simeq c_i^\star/2$: here free ceRNA concentration is roughly half the theoretical maximum; spontaneous ceRNAs decay and  miRNA-mediated degradation have similar weight.
\item[c.] If $[\mu]\gg\mu_{0,i}$ (or $\mu_{0,i}/[\mu]=\mathcal{O}(\epsilon)$), then $[m_i]\ll m_i^\star$ and $[c_i]\simeq c_i^\star$: here the levels of complexes are largest, while free ceRNAs are roughly absent; miRNA-mediated degradation is the prevailing channel of ceRNA decay.
\end{enumerate}
It is reasonable to expect that in cases a. and c. the steady state level of free ceRNA $[m_i]$ will only be weakly sensible to (small) variations in $[\mu]$. We shall call these regimes `Free' and `Bound' ($\mathcal{F}$and $\mathcal{B}$ for brevity), respectively. In case b., instead, the microRNA concentration lies in the dynamical range of $F_i$, so that $[m_i]$ will respond to (small) variations in $[\mu]$. We will the call this regime `Susceptible' ($\mathcal{S}$ for brevity). The outlook is that once the kinetic parameters of the ceRNAs are given, the $\mu_{0,i}$'s are given as well, and the miRNA level suffices to know whether a ceRNA is in the $\mathcal{F}$, $\mathcal{S}$ or $\mathcal{B}$ regime.

Biologically reasonable values of the model parameters ($k_i^+\sim 10^{-3} nM^{-1}s^{-1}$, $d_i\sim 10^{-4}s^{-1}$ as used in \cite{wang}) suggest that $\frac{d_i}{k_i^+}\sim\mu_{0,i}$ should approximatively have nanomolar order of magnitude, which is comparable to the range of miRNA concentrations \cite{concentration,Liang}.
 Note however that analysis of the RNAi enzyme complex has shown that its kinetics can vary substantially across different targets and that it is strongly affected by the degree of complementarity \cite{haley}. 

This means that, in principle, different targets may have very different $\mu_{0,i}$'s and may thus be located in distinct regimes of regulation at fixed $[\mu]$ and that the three states ($\mathcal{F,S,B}$) are actually assumed by ceRNAs. 
Furthermore, different miRNAs species may have concentrations spanning many orders of magnitude in a given cell type  \cite{mullokandov}.
 For instance, it has been experimentally demonstrated that only the most abundant miRNAs have significant impact on gene expression and mediate target suppression  \cite{mullokandov}, suggesting that ceRNAs are 'free' from miRNAs when their regulators have very low concentrations.
Moreover, it has been observed that protein production in presence of miRNA is highly repressed below a given threshold level for mRNA transcriptional activity and it responds sensitively to transcription above this threshold 
\cite{threshold}, suggesting the transition from an unexpressed, bound regime to an expressed, susceptible one.

To illustrate the emergent interactions between ceRNAs, we plot in Figure 2 the steady-state levels of $m_1$, $m_2$ and $\mu$ as a function of $b_1$ in a system with $N=2$ in which all other kinetic parameters are fixed.
\begin{figure}
\begin{center}
\includegraphics[width=8.5cm]{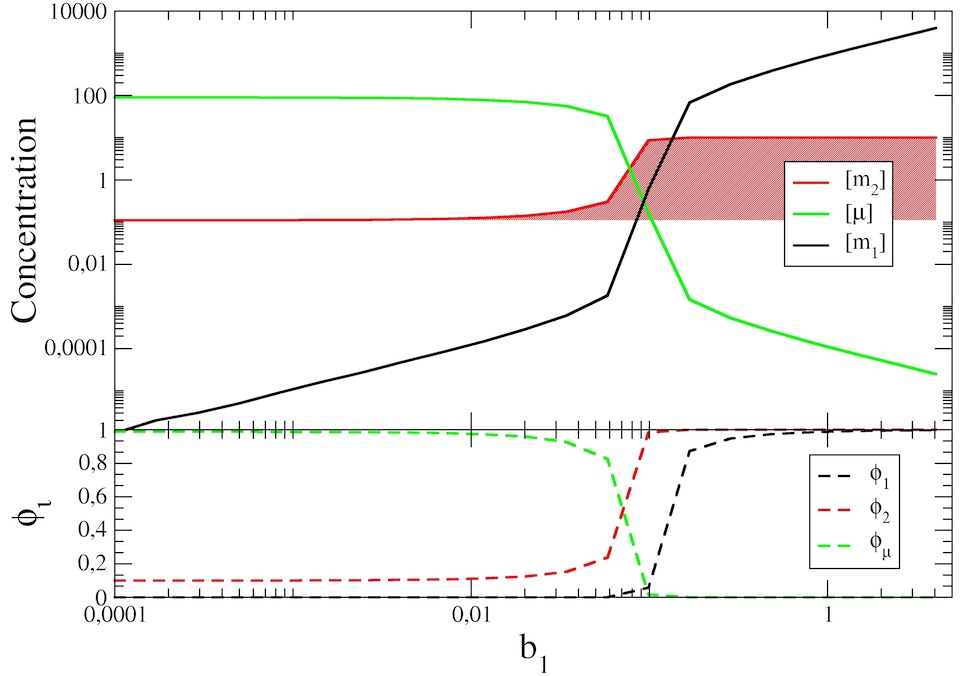}
\caption{{\bf Top}: steady state concentrations in a system with $N=2$ ceRNAs, obtained by fixing all parameters but the transcription rate $b_1$ of ceRNA 1.  {\bf Bottom}: fractions of free molecules, namely $\phi_1=[m_1]/([m_1]+[c_1])$, $\phi_2=[m_2]/([m_2]+[c_2])$, $\phi_\mu=[\mu]/([\mu]+[c_1]+[c_2])$, versus $b_1$. The dynamical range of the cross-talk interaction between the two ceRNAs corresponds to the window where free and bound molecules have similar concentration, i.e. to the $\mathcal{S}$ regime. Parameters for this case (in their respective units) are as follows: $b_2 = 10$, $d_1 = d_2 = \delta = 1$, $k_1^+ = 10^3$, $k_2^+ = 1$, $k_1^-= k_2^- = 10^{-3}$, $\sigma_1 = \sigma_2 = 10$, $\kappa_1 = \kappa_2 = 1$.}
\end{center}
\end{figure}
One clearly sees that, even within this basic model, a change in the transcription rate of a ceRNA can affect the steady state concentration of a different ceRNA.
The shaded area in the top panel of Figure 2 highlights the difference between the steady state level of ceRNA2  with or without (i.e. for $b_1=0$) its competitor $m_1$: $[m_2]$ is sensible to variations of  $b_1$ (via the change of the free miRNA concentration $[\mu]$, which is correspondingly dropping) only in an intermediate, narrow interval.

We will show that the intensity of such a 'cross-talk' depends on the regimes to which the ceRNAs belong. In essence, strong interactions can be achieved only (i) (symmetrically) between ceRNAs in the $\mathcal{S}$-regimes, and (ii) (asymmetrically) from a ceRNA in the $\mathcal{B}$-regime to a ceRNA in the $\mathcal{S}$-regime. This scenario will define a selective, possibly asymmetric channel of communication which links ceRNAs targeted by a common miRNA.

Our goal here is to characterize 'cross-talk interactions' quantitatively by computing the `susceptibilities' ($i,j=1,\ldots, N$)
\begin{equation}
\chi_{ij}=\frac{\partial [m_i]}{\partial b_j}~~,
\end{equation}
which for $i\neq j$ measure the magnitude of the interaction between ceRNA $m_i$ and ceRNA $m_j$. Using (\ref{sseq}) we get
\begin{equation}\label{chi}
\chi_{ij}=\frac{\partial}{\partial b_j}[m_i^\star F_i([\mu])]=\frac{F_i([\mu])}{d_i}\delta_{ij}+m_i^\star \frac{\partial F_i}{\partial[\mu]} \frac{\partial[\mu]}{\partial b_j}~~.
\end{equation}
The cross-susceptibility (the last term in (\ref{chi})) can thus be seen as the product of two factors: the response of the miRNA level to perturbations of the transcription rate of ceRNA $m_j$, and the response of the level of ceRNA $m_i$ ($[m_i]=m_i^\star F_i$) to perturbations of the miRNA level. For the latter we get
\begin{equation}\label{expansion2}
\frac{\partial F_i}{\partial [\mu]}=-\frac{\mu_{0,i}}{(\mu_{0,i}+[\mu])^2}\simeq 
\begin{cases}
-1/\mu_{0,i}& i\in \mathcal{F}\\
-1/(4\mu_{0,i})& i\in \mathcal{S}\\
-\mu_{0,i}/[\mu]^2 & i\in\mathcal{B}
\end{cases}
\end{equation}

Note that, expectedly, this function is negative and is largest for $[\mu]\simeq \mu_{0,i}$, i.e. when the ceRNA is in the $\mathcal{S}$-regime. In order to compute $\partial [\mu]/\partial b_j$, we need an explicit expression for $[\mu]$, which should be obtained from the steady state condition (\ref{sseq}). We re-write this as
\begin{equation}
\label{mu_eq0}
[\mu] \left[ \delta +\sum_i b_i z_i F_i([\mu])\right]=\beta~~,
\end{equation}
where $z_i=\sigma_i/[\mu_{0,i}(\sigma_i+\kappa_i)]$. Equation (\ref{mu_eq0}) tells us that if $\sigma_i=0$ for all $i$ (i.e. if complex decay is purely catalytic) then $[\mu]=\beta/\delta$ and no cross-talk is achievable since $[\mu]$ is independent of $b_j$. If however $\sigma_i\neq 0$ for some $i$, then other solutions are possible. In particular, (\ref{mu_eq0}) is an algebraic equation of order $N+1$ at most, an approximate solution of which can be obtained under the assumption that ceRNAs can be separated in the different regimes defined above. Using the fact that, up to next-to-leading order in $\epsilon\ll 1$,
\begin{equation}
\label{expansion}
F_i([\mu])\simeq 
\begin{cases}
1-[\mu]/\mu_{0,i}& i\in \mathcal{F}\\
\frac{1}{2}-([\mu]-\mu_{0,i})/(4\mu_{0,i})& i\in  \mathcal{S}\\
\mu_{0,i}/[\mu]& i\in \mathcal{B}
\end{cases}
\end{equation}
and neglecting (when necessary) terms of order $2$ or higher, one finds that see section {\it Derivation of miRNA steady-state concentration} of Supporting Text for a detailed derivation)
\begin{equation}\label{muu}
[\mu]\simeq\frac{\beta-\sum_{i\in\mathcal{B}}b_i w_i-\frac{1}{4}\sum_{i\in \mathcal{S}} b_i w_i}{\delta+\sum_{i\in\mathcal{F}} b_i z_i+\frac{1}{4}\sum_{i\in \mathcal{S}} b_i z_i}~~.
\end{equation}
One now sees that
\begin{equation}\label{chi1}
\chi_{\mu j}\equiv\frac{\partial[\mu]}{\partial b_j}\simeq  -w_j \,\chi_{\mu\mu} \times
\begin{cases}
[\mu]/\mu_{0,j}& j\in \mathcal{F}\\
(\mu_{0,j}+[\mu])/(4\mu_{0,j}) & j\in\mathcal{S}\\
1 & j\in\mathcal{B}
\end{cases}
\end{equation}
where we have defined the shorthand
\begin{equation}
\chi_{\mu\mu}\equiv\frac{\partial [\mu]}{\partial \beta}=\left(\delta+\sum_{i\in\mathcal{F}} b_i z_i+\frac{1}{4}\sum_{i\in \mathcal{S}} b_i z_i\right)^{-1}
\end{equation} 
Given a shift in the level of a ceRNA, the response of the miRNA is always negative (since an increase in $b_j$ causes an increase in the level of complexes $c_j$). Also, if $m_j\in \mathcal{F}$ then $[\mu]/\mu_{0,j}\ll 1$, i.e. the level of microRNA is roughly insensitive to small changes of the production rate of ceRNAs in the Free regime. 

Finally, combining (\ref{expansion2}) and (\ref{chi1}) we obtain for $\chi_{ij}$ 
\begin{equation}\label{chitot}
\chi_{ij}=\frac{1}{d_i}\left[F_i([\mu])\delta_{ij}+\frac{b_i w_j\, \chi_{\mu\mu}}{4[\mu]}W_{R(i),R(j)}\right]
\end{equation}
where $W_{R(i),R(j)}$ is a coefficient that depends only on the regimes $R(i)$ and $R(j)$ to which $i$ and $j$ belong. In other words, $R(i),R(j)\in\{\mathcal{F},\mathcal{S},\mathcal{B}\}$ and the $3\times 3$ matrix $\widehat{W}$ is given by
\begin{multline}\label{matra}
\widehat{W}=\left(
\begin{aligned}
4\frac{[\mu]^2}{\mu_{0i}\mu_{0j}}\quad \frac{[\mu]}{\mu_{0i}}\frac{\mu_{0j}+[\mu]}{\mu_{0j}}\quad 4\frac{[\mu]}{\mu_{0i}}\\
\frac{[\mu]^2}{\mu_{0i}\mu_{0j}}\quad \frac{[\mu]}{\mu_{0i}}\frac{\mu_{0j}+[\mu]}{4\mu_{0j}}\quad \frac{[\mu]}{\mu_{0i}}\\
4\frac{\mu_{0i}}{\mu_{0j}}\quad \frac{\mu_{0i}}{[\mu]}\frac{\mu_{0j}+[\mu]}{\mu_{0j}}\quad 4\frac{\mu_{0i}}{[\mu]}\\
\end{aligned}
\right)=
\\
=\left(
\begin{aligned}
\mathcal{O}(\epsilon^2)\quad \mathcal{O}(\epsilon) \quad \mathcal{O}(\epsilon) \\
\mathcal{O}(\epsilon)\quad \mathcal{O}(1) \quad \mathcal{O}(1) \\
\mathcal{O}(\epsilon^2)\quad \mathcal{O}(\epsilon) \quad \mathcal{O}(\epsilon)
\end{aligned}
\right)
\end{multline}
Three important observations can now be made about $\chi_{ij}$. First, since both terms (\ref{expansion2}) and (\ref{chi1})  are negative, the cross-talk between ceRNAs tends to correlate their levels. Second, the matrix $\widehat{W}$ is not symmetric, as might perhaps have been expected. Finally, for $i\neq j$ all of the elements of $\widehat{W}$ are of order $\epsilon$ or smaller except $W_{\mathcal{S},\mathcal{S}}$ and $W_{\mathcal{S},\mathcal{B}}$, which are of order 1. This implies that, in this scenario, two types of effective interactions arise: the first one encodes the response of a ceRNA in the $\mathcal{S}$-regime to a perturbation of another ceRNA in the $\mathcal{S}$-regime, and it is symmetric; the second one encodes the response of a ceRNA in the  $\mathcal{S}$-regime to a perturbation of a ceRNA in the $\mathcal{B}$-regime, and it is not symmetric (i.e. perturbing the `susceptible' ceRNA the `bound' one will not respond). 

In words, the scenario described here corresponds to a linear response theory in which a change in the transcription rate of a ceRNA (i.e. $b_j\to b_j+\delta b_j$) induces a shift in $[m_j]$ (i.e. $[m_j]\to [m_j]+\chi_{jj}\delta b_j$ with $\chi_{jj}>0$) and a shift in the level of miRNA (i.e. $[\mu]\to[\mu]+\chi_{\mu j}\delta b_j$ with $\chi_{\mu j}<0$). In turn, this affects $[m_i]$ (i.e. $[m_i]\to [m_i]+\chi_{ij}\delta b_j$ with $\chi_{ij}>0$). So for instance if $\delta b_j>0$ then $[m_j]$ increases, $[\mu]$ decreases and $[m_i]$ increases. The quantities $\mu_{0,i}\propto 1/k_i^+$ can be seen to induce a hierarchy of interactions:  ceRNAs in the $\mathcal{B}$-regime (higher binding affinity) can unidirectionally affect ceRNAs in the $\mathcal{S}$-regime, which in turn may influence other ceRNAs in the $\mathcal{S}$-regime. On the other hand, ceRNAs in the $\mathcal{F}$ regime (lower binding affinity) interact weakly with the rest of the system and fluctuations in their transcription rates do not propagate to other ceRNAs. It is important to remark that cross-talk appears only when the miRNA level is in a specific range, implying that the structure of the emergent interaction network is flexible and dynamical: the set of ceRNA species that interact may change upon varying $[\mu]$.

The emergence of selectivity and directionality as features of the cross-talk can be seen in a concrete case in Figure 3, where we plot the susceptibilities for a system of $N=4$ ceRNAs. 
\begin{figure}
\begin{center}
\includegraphics[width=8.5cm]{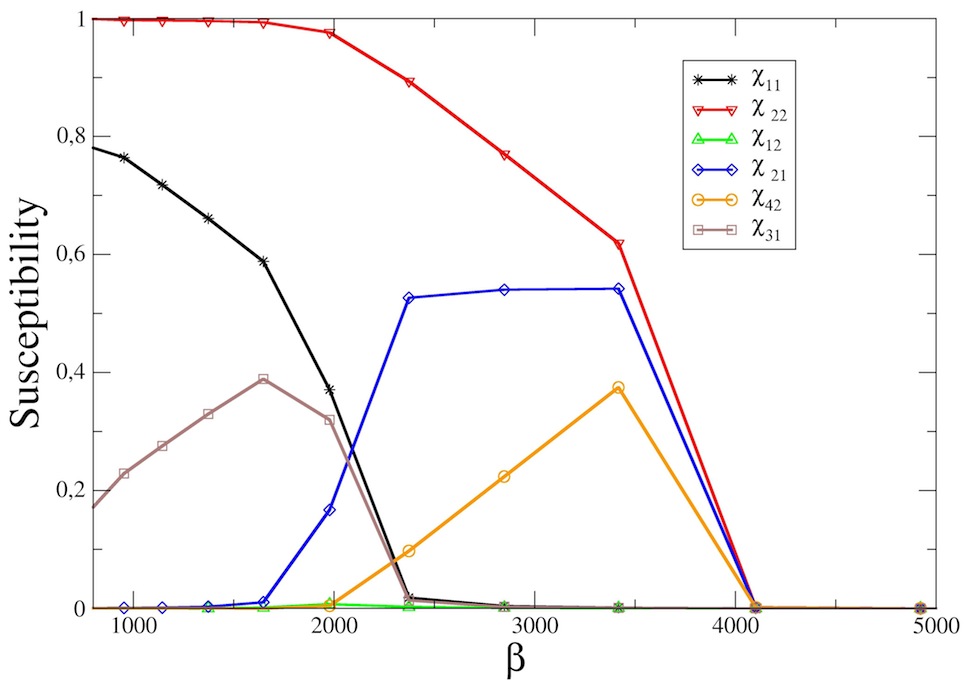}
\caption{Susceptibilities $\chi_{ij}$ in a system of $N=4$ ceRNAs, as a function of miRNA transcription rate $\beta$ (all other parameters being fixed). In this example ceRNAs are cast in two groups: group A, formed by ceRNAs $m_1$ and $m_3$, and group B, formed by ceRNAs $m_2$ and $m_4$. CeRNAs belonging to the same group share identical kinetic parameters. In particular, $\mu_{0,1}=\mu_{0,3}\ll\mu_{0,2}=\mu_{0,4}$. 
 While for $\beta$ smaller than about 500 no cross-talk is observed, as $\beta$ increases a symmetric interaction between ceRNAs in group A (of magnitude comparable to the self-susceptibilities) appears:  $\chi_{31}=\chi_{13} \simeq \chi_{11}=\chi_{33}$. As $\beta$ increases further, this interaction is switched off, and ceRNAs in group B begin to cross-talk instead: $\chi_{42}=\chi_{24}\simeq \chi_{22}=\chi_{44}$.
In this region, a change of transcription of a ceRNA in group A can affect the level of ceRNAs in group B, but not viceversa (asymmetric cross-talk): $\chi_{21}=\chi_{23}=\chi_{41}=\chi_{43}>>\chi_{12}=\chi_{32}=\chi_{14}=\chi_{34}$. Finally, for sufficiently large $\beta$ all no cross-talk takes place.}
\end{center}
\end{figure}
One sees that different interactions are switched on in different ranges of values for  the miRNA transcription rate, leading to a gradual modification of the structure of the interaction network as $\beta$ changes. Notice that heterogeneity in the quantities $\mu_{0,i}$ leads to interaction asymmetry. A schematic summary of the cross-talk in this system is given in Figure 4.
\begin{figure*}
\begin{center}
\includegraphics[width=17cm]{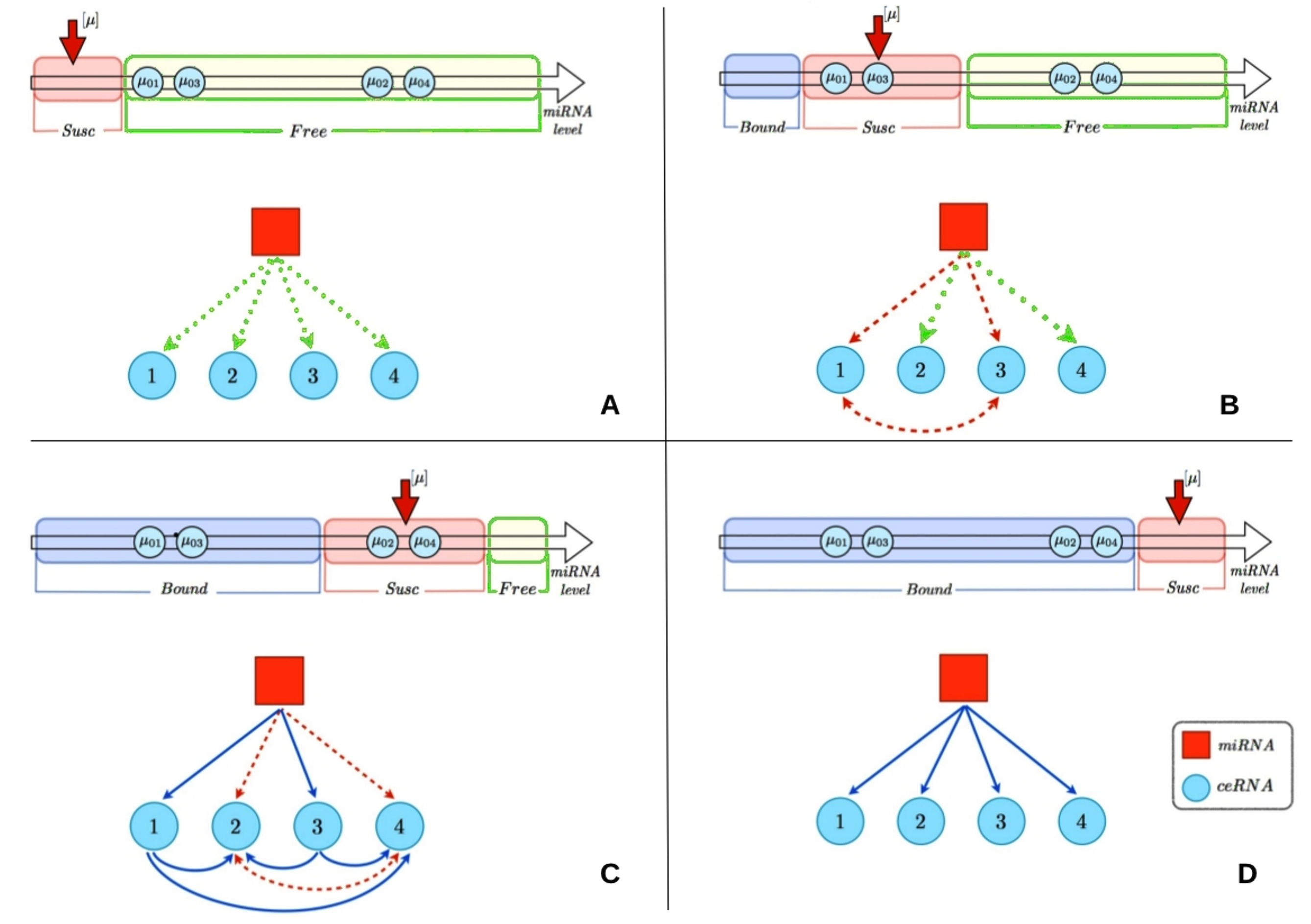}
\caption{Schematic representation of the patterns of interactions arising in a system of $N=4$ ceRNAs at different miRNA levels (increasing from A to D). {\bf A)} All ceRNAs are in the $\mathcal{F}$ regime ($[\mu]\ll\mu_{0,i}~\forall i$)  and there is no interaction between them.  {\bf B)} ceRNAs 1 and 3 are in the $\mathcal{S}$ regime ($[\mu]\simeq \mu_{0,1}\simeq \mu_{0,3}$) and a symmetrical interaction between them is switched on. {\bf C)} ceRNAs 1 and 3 are now in the $\mathcal{B}$ regime ($[\mu]\gg \mu_{0,1}\simeq \mu_{0,3}$) while ceRNAs 2 and 4 are in the $\mathcal{S}$ regime ($[\mu]\simeq \mu_{0,2}\simeq \mu_{0,4}$): the resulting interactions are symmetric between 2 and 4 and asymmetric of ceRNAs 1 and 3 on ceRNAs 2 and 4. {\bf D)} All ceRNAs are in the $\mathcal{B}$ regime ($[\mu]\gg\mu_{0,i}~ \forall i$)  and no cross-talk occurs.}
\end{center}
\end{figure*}

We notice that the intensity of the cross-talk described by (\ref{chitot}) is modulated by the factor
\begin{equation}\label{kjh}
\frac{b_i w_j\, \chi_{\mu\mu}}{4[\mu]} \simeq \frac{b_i w_j}{ 4 [\mu]\delta + \sum_{k\in \mathcal{S}} b_k w_k}
\end{equation}
where we used $[\mu]\simeq \mu_{0,k}$ for $k\in \mathcal{S}$ and neglected the contribution to $[\mu]$ due to ceRNAs in the $\mathcal{F}$-regime (see (\ref{muu})), which is of order $\epsilon$. Again, we see that an effective interaction requires some degree of stoichiometric degradation: if $w_j=0$ (i.e. if the complex decays in a purely catalytic manner) the cross-susceptibility vanishes. In addition, (\ref{kjh}) suggests that the magnitude of the interaction is weakened only by ceRNAs lying in the $\mathcal{S}$-regime, so that, even in presence of a large number of interacting ceRNA species, cross-talk can be large if the overall population of ceRNAs in the $\mathcal{S}$-regime is restricted. On the other hand, many factors, such as the overall population of ceRNAs in the $\mathcal{B}$-regime, affect the rate of miRNA transcription required in order for  a given ceRNA to be susceptible. 

A rough approximate expression for the range $\Delta \beta$ of values of the miRNA transcription rate where ceRNA $m_i$ is most responsive to the miRNA can be derived considering as `susceptible' a window $\Delta_\mu$ of $[\mu]$ values such that $\mu_{0,i}/2<[\mu]<3\mu_{0,i}/2$. If so, then
 \begin{equation}
\Delta \beta \simeq \frac{\Delta_\mu}{\chi_{\mu\mu}} \simeq \frac{\mu_{0,i}}{\chi_{\mu\mu}} \simeq \left(\mu_{0,i}\delta + \sum_{k\in \mathcal{S}} b_k w_k\right) ~~,
\end{equation}
where we used $\Delta_\mu\simeq \mu_{0,i}$. One sees that $\Delta\beta$ mainly depends on the transcription rate of the overall population of ceRNAs in the susceptible regime and on the degree of stoichiometricity of degradation.

Finally, we observe that an analogous cross-talk scenario emerges for a system in which $M$ miRNA species share the same target RNA: the level of a miRNA species may be highly susceptible to a change in the transctiption rate of a different miRNA when the level of the target RNA lies in a specific window (for details see section {\it The mirror system: one target, M miRNA species} in the Supporting Text).

\subsection*{$N$ ceRNA species, $M$ miRNA species}

Let us now consider the general case of a system formed by $M$ miRNA species $\mu_\alpha$ ($\alpha=1,\ldots,M$) and $N$ ceRNA species $m_i$ ($i=1,\ldots,N$), defined by the rates
\begin{gather*}
\emptyset \xrightleftharpoons[d_i]{b_i} m_i ~~~~~~~~~~~~
\emptyset \xrightleftharpoons[\delta_\alpha]{\beta_\alpha} \mu_\alpha ~~~~~~~~~~~~
\mu_\alpha+m_i \xrightleftharpoons[k_{i\alpha}^-]{k_{i\alpha}^+} c_{i\alpha}\\
c_{i\alpha} \xrightharpoonup{\sigma_{i\alpha}} \emptyset ~~~~~~~~~~~~
c_{i\alpha} \xrightharpoonup{\kappa_{i\alpha}} \mu_\alpha 
\end{gather*}
and for which the following steady-equations hold:
\begin{gather}
[m_i]=\frac{b_i+\sum_\alpha k_{i\alpha}^-[c_{i\alpha}]}{d_i+\sum_\alpha k_{i\alpha}^+[\mu_\alpha]}\nonumber\\
[\mu_\alpha]=\frac{\beta_\alpha+\sum_i(k_{i\alpha}^-+\kappa_{i\alpha})[c_{i\alpha}]}{\delta_\alpha+\sum_i k_{i\alpha}^+[m_i]}\label{ssse}\\
[c_{i\alpha}]=\frac{ k_{i\alpha}^+[\mu_\alpha][ m_i]}{\sigma_{i\alpha}+k_{i\alpha}^-+\kappa_{i\alpha}}\nonumber 
\end{gather}
Again, we shall focus on computing ceRNA sensitivities to perturbations of transcription rates of other ceRNAs. Neglecting higher order interactions involving two or more miRNAs (which is justified for a large, sparse miRNA-ceRNA network in absence of connectivity correlations) we have
\begin{equation}\label{cchi}
\chi_{ij}=\frac{\partial [m_i]}{\partial b_j}\simeq \sum_{\alpha}\frac{\partial [m_i]}{\partial [\mu_\alpha]}\frac{\partial [\mu_{\alpha}]}{\partial b_j}\equiv \sum_\alpha\chi_{ij,\alpha}
\end{equation}
The $N M+N+M$ steady state equations (\ref{ssse}) can be reduced to $N+M$ coupled equations for the $N+M$ unknown $\{\mu_\alpha,m_i\}$ by eliminating the complexes. After some straightfoward algebra, we get $[m_i]=m_i^\star F_i(\{[\mu_\alpha]\})$ and $\mu_{\alpha}=\mu_{\alpha}^\star F_{\alpha}(\{[m_i]\})$
with
\begin{equation}
F_i=\left( 1+\sum_\alpha \frac{[\mu_\alpha]}{\mu_{0,i\alpha}}\right)^{-1}~~~~,~~~~
F_\alpha=\left( 1+\sum_i \frac{[m_i]}{m_{0,i \alpha}}\right)^{-1}~~,
\end{equation}
where 
\begin{gather}
\mu_{0,i\alpha}=\frac{d_i}{k^+_{i\alpha}}(1+\phi_{i\alpha})~~~~,~~~~
m_{0,i\alpha}=\frac{\delta_\alpha}{k^+_{i\alpha}}(1+\psi_{i\alpha})\\
\phi_{i\alpha}=\frac{k^-_{i\alpha}}{\sigma_{i\alpha}+\kappa_{i\alpha}}~~~~,~~~~
\psi_{i\alpha}=\frac{k^-_{i\alpha}+\kappa_{i\alpha}}{\sigma_{i\alpha}}
\end{gather}
In turn, the levels of free miRNA and ceRNA are described by
\begin{gather}
[\mu_\alpha] \left[ \delta_\alpha+\sum_i b_i z_{i\alpha} F_i\right]=\beta_\alpha\\
[m_i] \left[ d_i+\sum_\alpha \beta_\alpha \zeta_{i\alpha} F_\alpha\right]=b_i
\end{gather}
with $z_{i\alpha}=\sigma_{i\alpha}/[\mu_{0,i\alpha}(\sigma_{i\alpha}+\kappa_{i\alpha})]$ and $\zeta_{i\alpha}=(\sigma_{i\alpha}+\kappa_{i\alpha})/(m_{0,i\alpha}\sigma_{i\alpha})$. The quantity $F_i$ can be re-written as
\begin{equation}
\label{zi}
F_i=\frac{1}{Z_{i}^{(\alpha)}}\frac{1}{1+[\mu_\alpha]/\tilde{\mu}_{0,i\alpha}}
\end{equation}
where $Z_{i}^{(\alpha)}=1+\sum_{\gamma\neq\alpha}[\mu_\gamma]/\mu_{0,i\gamma}$
and $\tilde{\mu}_{0,i\alpha}=\mu_{0,i\alpha}Z_{i}^{(\alpha)}$. Note that sum in $Z_{i}^{(\alpha)}$ includes all miRNA species except for $\mu_\alpha$. 

Equation (\ref{zi}) tells us that, in presence of many miRNA species, we may account for the effect of species $\mu_\alpha$ on ceRNA $m_i$ by just re-scaling $F_i$ by $Z_i^{(\alpha)}$ and shifting the reference level $\mu_{0,i\alpha}$ by $Z_{i}^{(\alpha)}$. 
A simple interpretation of the above expressions can be gained by introducing an effective decay rate  $d_i^{(\alpha)}=d_i Z_{i}^{(\alpha)}$ and noting that  
\begin{equation}
[m_i]=
\frac{b_i}{d_i^{(\alpha)}} \frac{\tilde\mu_{0,i\alpha}}{[\mu]+\tilde\mu_{0,i\alpha}}~~~,~~~\tilde \mu_{0,i\alpha}=\frac{d_i^{(\alpha)}}{k^+_{i\alpha}}(1+\phi_{i\alpha})~~.
\end{equation}
One immediately recognizes the same form of the steady state equation (\ref{tre}) for the case $M=1$, and sees that $Z_{i}^{(\alpha)}$ ultimately plays the role of a factor accelerating the effective turnover. Note that all miRNAs targeting ceRNA $i$ give positive contributions to the sum $Z_{i}^{(\alpha)}$ and thus increase the effective turnover, but the most important contributions come from those miRNAs whose level $[\mu_{\gamma}]$ is high respect to the term $\mu_{0i,\gamma}$. 

By analogy with the case $M=1$, we will say that a ceRNA is free with respect to miRNA $\mu_\alpha$ (and write $i\in\mathcal{F}(\alpha)$) if $[\mu_\alpha]\ll\tilde\mu_{0,i\alpha}$; it will be `susceptible' with respect to $\mu_\alpha$ (or $i\in\mathcal{S}(\alpha)$) if $[\mu_\alpha]\simeq \tilde\mu_{0,i\alpha}$; it will be `bound' with respect to $\mu_\alpha$ (or $i\in\mathcal{B}(\alpha)$) if $[\mu_\alpha]\gg\tilde\mu_{0,i\alpha}$. Note that being bound with respect to a miRNA species is sufficient for a ceRNA to be translationally repressed. For consistency, a ceRNA can only be bound with respect to one miRNA species (in that case it will be free with respect to all other miRNAs). Separating the different regimes we have
\begin{equation}
\label{expansion3}
F_i\simeq 
\begin{cases}
[Z_{i}^{(\alpha)}]^{-1}(1-[\mu_\alpha]/\tilde\mu_{0,i\alpha})& i\in \mathcal{F}(\alpha)\\
\frac{1}{2}[Z_{i}^{(\alpha)}]^{-1}[1-(\mu-\tilde\mu_{0,i\alpha})/(2\tilde\mu_{0,i\alpha})]& i\in\mathcal{S}(\alpha)\\
\tilde \mu_{0,i\alpha}/[\mu_\alpha]& i\in \mathcal{B}(\alpha)
\end{cases}
\end{equation}
In turn, for the levels of free miRNAs we obtain
\begin{equation}
[\mu_\alpha]\simeq\frac{\beta_\alpha-\sum_{i\in \mathcal{B}(\alpha)} b_i  w_{i\alpha} 
-\frac{1}{4}\sum_{i\in \mathcal{S}(\alpha)} b_i w_{i\alpha}}{\delta_\alpha+
\sum_{i\in \mathcal{F}(\alpha)} b_i \tilde z_{i\alpha}+
\frac{1}{4}\sum_{i\in \mathcal{S}(\alpha)} b_i \tilde z_{i\alpha}}
\end{equation}
where $w_{i\alpha}=\sigma_{i\alpha}/(\sigma_{i\alpha}+\kappa_{i\alpha})$ and $\tilde z_{i\alpha}=\sigma_{i\alpha}/[\tilde\mu_{0,i\alpha}(\sigma_{i\alpha}+\kappa_{i\alpha})]$.
One may now compute the different terms of the susceptibilities. For the quantity $\chi_{ij,\alpha}$ (see (\ref{cchi})) we finally get the $M>1$ analog of (\ref{chitot}), i.e. 
\begin{equation}\label{chitot_NM}
 \chi_{ij,\alpha}=\frac{1}{d_i }\left[F_i\delta_{ij}+  \frac{b_i \, \tilde w_{j\alpha} \, \chi_{\alpha \alpha}}{4 \, Z_i^{(\alpha)} \, [\mu_\alpha] \, } \, W_{\alpha;R(i),R(j)}\right]
\end{equation}
where the matrices $\widehat{W}_\alpha$ are given by
\begin{equation}
\widehat{W}_\alpha=\left(
\begin{aligned}
4\frac{[\mu_{\alpha}]^2}{\tilde\mu_{0,i\alpha}\tilde\mu_{0,j\alpha}}\quad 
\frac{[\mu_{\alpha}]}{\tilde\mu_{0,i\alpha}}\frac{\tilde\mu_{0,j\alpha}+[\mu_\alpha]}{\tilde\mu_{0,j\alpha}}\quad 4\frac{[\mu_{\alpha}]}{\tilde\mu_{0,i\alpha}}\\
\frac{[\mu_{\alpha}]^2}{\tilde\mu_{0,i\alpha}\tilde\mu_{0,j\alpha}}\quad 
\frac{[\mu_\alpha]}{\tilde\mu_{0,i\alpha}}\frac{\tilde\mu_{0,j\alpha}+[\mu_\alpha]}{4\tilde\mu_{0,j\alpha}}\quad \frac{[\mu_\alpha]}{\tilde\mu_{0,i\alpha}}\\
4\frac{\tilde\mu_{0,i\alpha}}{\tilde\mu_{0,j\alpha}}\quad 
\frac{\tilde\mu_{0,i\alpha}}{[\mu_{\alpha}]}\frac{\tilde\mu_{0,j\alpha}+[\mu_\alpha]}{\tilde\mu_{0,j\alpha}}\quad 4\frac{\tilde\mu_{0,i\alpha}}{[\mu_{\alpha}]}\\
\end{aligned}
\right)
\end{equation}
In this case, the intensity of the cross-talk described by (\ref{chitot_NM}) is modulated by the factor
\begin{equation}
\frac{b_i w_{j\alpha}\, \chi_{\alpha\alpha}}{4Z_i^{(\alpha)} [\mu_\alpha]} \simeq \frac{b_i w_j}{ Z_i^{(\alpha)}(4 [\mu_\alpha]\delta_\alpha + \sum_{k\in \mathcal{S}(\alpha)} b_k w_{k\alpha})}~~.
\end{equation}
We therefore conclude that miRNA $\mu_\alpha$ gives a relevant contributions to the overall susceptibility $\chi_{ij}$ if either:
\begin{enumerate}
\item[(i)] $i\in\mathcal{S}(\alpha)$;
\item[(ii)]  $j\in\mathcal{S}(\alpha)$ or $j\in\mathcal{B}(\alpha)$ (in the latter case, $\mu_\alpha$ is the main repressor of $m_j$);
\item[(iii)] $Z_{i}^{(\alpha)}\simeq 1$, i.e. ceRNA $m_i$ has few repressors besides $\mu_\alpha$;
\item[(iv)] few ceRNA species belong to $\mathcal{S}(\alpha)$, so that dilution is limited.
\end{enumerate}
In summary, the effect of `background' miRNAs which do not mediate interactions is an increase in the effective rate of decay, and consequently a shift in the susceptibility threshold. On the other hand, the effect of background ceRNAs is a dilution of the cross-talk among ceRNAs, as seen in the case $M=1$. An illustrative example of the interactions arising among ceRNAs in a system with $N=4$ and $M=3$ is shown in Figure 5.
\begin{figure}
\begin{center}
\includegraphics[width=8.5cm]{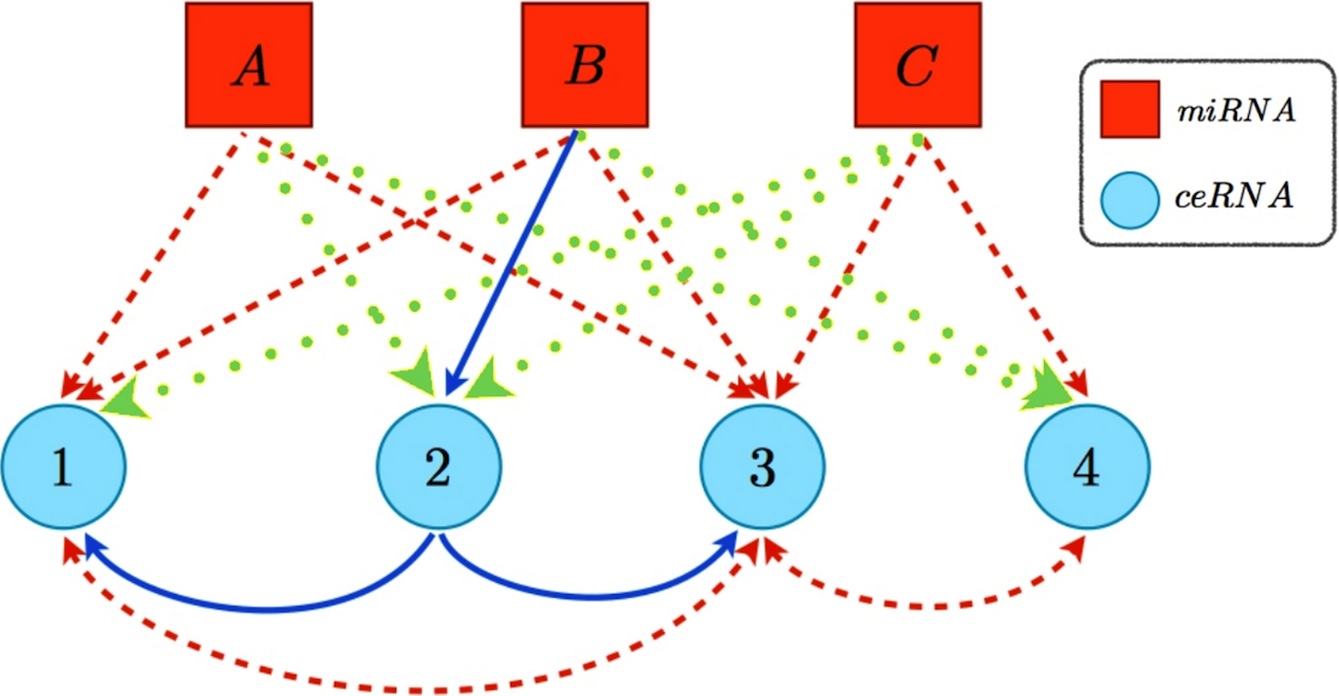}
\caption{Schematic representation of a system of $N=4$ ceRNAs species and $M=3$ miRNA species. Continuous blue  arrows link miRNA $\alpha$ to ceRNA $i$ if $i\,\in \mathcal{B}(\alpha)$; dashed red arrows are found if $i\,\in \mathcal{S}(\alpha)$; dotted green arrows are found if $i\,\in \mathcal{F}(\alpha)$. In this case, ceRNAs $1$ and $3$ are both in $\mathcal{S}(A)$ and  $ \mathcal{S}(B)$, ceRNAs $3$ and $4$ are in $\mathcal{S}(C)$, and ceRNA is $1$ in $\mathcal{B}(A)$.  This situation results in the following interactions: symmetric cross-talk between ceRNAs $1$ and $3$, mediated by miRNAs $A$ and $B$;  symmetric cross-talk between ceRNAs $3$ and $4$, mediated by miRNA $C$;  asymmetric cross-talk from ceRNA $2$ to ceRNA $1$, mediated by miRNA $B$;  asymmetric cross-talk from ceRNA $2$ to ceRNA $3$, mediated  by miRNA $B$.}
\end{center}
\end{figure}

To conclude, we notice that network topology can play an important role as interaction enhancer. For instance (see section {\it The role of topology} in the Supporting Text),  cross-talk can take place among ceRNAs in the Free regime (in spite of the small $\chi_{ij,\alpha}$) provided they are commonly targeted by a large number of miRNA species. In other terms: interactions between ceRNAs can be mediated by a large number of miRNA species which individually would only weakly dampen ceRNA levels. However, in order to achieve efficient cross-talk strong correlations in the network connectivity are needed, so that highly clustered networks can allow for much stronger cross-talk than random graphs.

\subsection*{Steady state fluctuations}

Genetic circuits that regulate cellular functions are subject to stochastic fluctuations, specifically in the levels of the different molecular species that interact \cite{noise,noise2}. Noise, far from being just a nuisance, plays an essential role in cellular activities, for example by enabling coordination of gene expression across large regulons, or by allowing for probabilistic differentiation of otherwise identical cells \cite{noise4}. On the other side, a noisy gene expression is potentially harmuful in many situations: in developmental circuits, for example, it can lead either to arrested development, aberrant positional expression of tissue specific genes or over-representation of specific cell types \cite{noise5}. If there is only a relatively narrow protein level which is optimal, some sort of tuning must act to prevent fluctuations outside the functional range. Cross-talk of the type discussed so far may either result in an  amplification of upstream fluctuations or represent an efficient noise buffering mechanism. To analyze this issue in some detail, we focus on the role of transcriptional noise, the primary cause of variability in gene expression among cells in isogenic populations \cite{noise3}. If one assumes that extrinsic transcriptional noise is the dominant source of stochasticity and neglects molecular noise entirely, it is possible to estimate concentration fluctuations in the ceRNA-miRNA networks at steady state, obtaining expressions valid in the linear response regime.

Let us consider for simplicity a system of $N=2$ ceRNA species and $M=1$ miRNA species, and let $P(\mathbf{r})$  denote a distribution of transcription rates (where $\mathbf{r}=\{b_1,b_2,\beta\}$), such that an ensemble of systems at steady state can be constructed by sampling a vector $\mathbf{r}$ from $P(\mathbf{r})$ for each system in the ensemble. For $P(\mathbf{r})$ one may for simplicity take a Gaussian, i.e.
\begin{equation}
P(\mathbf{r})\propto \exp\left[
-\frac{1}{2}(\mathbf{r}-\overline{\mathbf{r}})^T \Sigma^{-1} (\mathbf{r}-\overline{\mathbf{r}})\right]
\end{equation}
where $\overline{\mathbf{r}}=\{\overline{b}_1,\overline{b}_2,\overline{\beta}\}$ is the mean and $\Sigma$ is the correlation matrix of inputs. Clearly, a distribution of transcription rates induces a distribution of steady state concentrations. The latter is what we aim at characterizing.

If variability in transcription rates is sufficiently small, we can expand the steady state levels $\boldsymbol{\ell}= \{[m_1],[m_2],[\mu]\}$ (note that $\boldsymbol{\ell}\equiv\boldsymbol{\ell}(\mathbf{r})$) around $\overline{\mathbf{r}}$ (small noise expansion), obtaining
\begin{equation}\label{pob}
\ell_i \simeq \overline{\ell_i}+\sum_k \chi_{ik}(r_k-\overline{r}_k)~~~~~,~~~~~\chi_{ik}=\frac{\partial \ell_i}{\partial r_k}~~,
\end{equation}
where $\overline{\ell_i}\equiv \ell_i(\overline{\mathbf{r}})$. In this approximation:
\begin{multline}
\text{Prob}(\boldsymbol{\ell}=\mathbf{x})=\int P(\mathbf{r}) \delta[\boldsymbol{\ell}(\mathbf{r})-\mathbf{x}] \text{d}\mathbf{r} = \\
=N \exp\left[-\frac{1}{2}(\boldsymbol{\ell}-\overline{\boldsymbol{\ell}})^T
X (\boldsymbol{\ell}-\overline{\boldsymbol{\ell}}) \right]
\end{multline}
where $X=(\hat\chi^{-1})^T \Sigma^{-1}\, \hat\chi^{-1}$, $\hat\chi$ being the matrix of susceptibilities defined in (\ref{pob}). The joint probability distribution and the susceptibility matrix can then be used to characterize steady state fluctuations and correlations, e.g.
\begin{equation} 
\label{sigma}
 \sigma^2_{i}\equiv\overline{(\ell_i- \overline{\ell_i})^2}= \sum_{j,k} \chi_{ij}\,\chi_{ik}\,\Sigma_{jk}
 \end{equation}
For uncorrelated transcription rates the covariance matrix $\Sigma$ is diagonal and (\ref{sigma}) reduces to $\sigma^2_{i}=\sum_{k} \chi_{ik}^2\, \Sigma_{kk}$: expectedly, each term positively contributes to increase the noise. 
As shown in Figure 6, fluctuations can become very large in the susceptible regime as the system is strongly coupled, possibly limiting the efficiency of signaling in ceRNA network .
\begin{figure}
\begin{center}
\includegraphics[width=8.5cm]{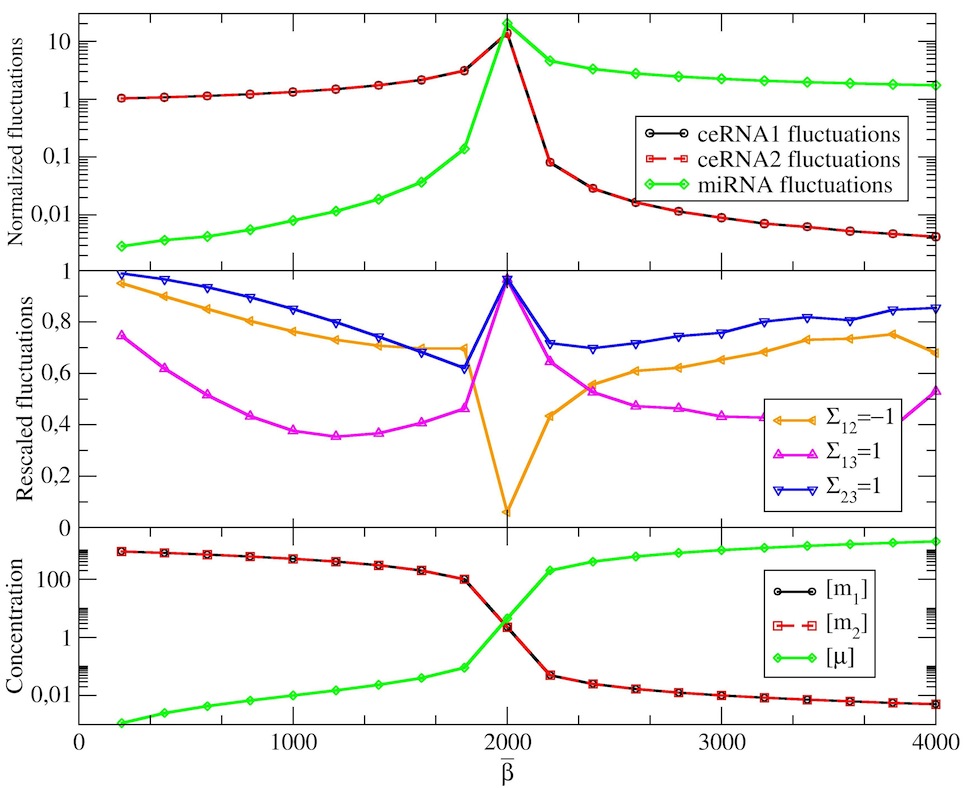}
\caption{Level fluctuations induced by transcriptional noise in a system with $N=2$ ceRNA species and $M=1$ miRNA species. We took Gaussian distributions for the transcription rates $\beta$, $b_1$ and $b_2$, keeping the ratio between the average and the width fixed in each case. All parameters and distributions of rates are also kept fixed, except for the miRNA transcription rate distribution $P(\beta)$, which is parametrized by its average value $\overline{\beta}$. In this case: $\overline{b_1}= \overline{b_2} = 10^{3}$,  $d_1 = d_2 = \delta = 1$, $k_1^+=k_2^+ = 10^{2}$, $k_1^-= k_2^- = 0$, $\sigma_1 = \sigma_2 = 10$, $\kappa_1 = \kappa_2 =0$.  Top: normalized fluctuations (ratio between the width of the fluctuations in the interacting and the non-interacting system, the latter corresponding to $k_1^+=k_2^+=0$) for uncorrelated distributions of transcription rates. Center: ratio between the normalized fluctuations  of the level of ceRNA 1 obtained in presence of correlations and in the uncorrelated case. Yellow line: maximal anti-correlation between $b_1$ and $b_2$ ($\Sigma_{12}=-1$); purple line: maximal correlation between $b_1$ and $\beta$ ($\Sigma_{13}=1$); blue line: maximal correlation between $b_2$ and $\beta$ ($\Sigma_{23}=1$)). Bottom: average molecular levels.}
\end{center}
\end{figure}

In presence of correlations at the transcriptional level, however, the signs of off-diagonal terms become crucial. Recalling that, generically, $\chi_{i,\mu}<0$, $\chi_{i,j}>0$, and $\chi_{\mu,i}<0$, one sees that anti-correlated ceRNA transcriptions and correlated miRNA-ceRNAs transcriptions may lead to a reduction of fluctuations with respect to the uncorrelated case, as shown again in Figure 6. On the other hand, negative miRNA-ceRNA correlations and positive ceRNA-ceRNA correlations strongly amplify fluctuations. In other terms, miRNA-mediated cross-talk coupled with correlation of transcriptional inputs may represent a powerful noise processing mechanism.

\subsection*{Detection of miRNA-mediated cross-talk from gene expression data}

A key issue of the ceRNA scenario concerns the detection of cross-talk in gene expression data, typically from correlations or related quantities.  It is important to note that, within the theoretical framework we discuss, the presence of statistically significant correlations between ceRNAs is not necessarily a signature in this sense. Indeed, the Pearson correlation coefficient between ceRNAs, for independent transcription rates, reads
\begin{equation} 
\label{pearson}
 \rho_{12}=\frac{\sum_k \chi_{1k}\,\chi_{2k}\,\sigma^2_k}{\sqrt{(\sum_k \chi_{1k}^2 \sigma_k^2)(\sum_k \chi_{2k}^2 \sigma_k^2)}}~~,
 \end{equation}
 where $\sigma^2_k$ is the variance of $r_k$. However if $\chi_{12}=0$ then
\begin{equation} 
 \rho_{12} = A\, \frac{\partial [m_1]}{\partial\beta} \, \frac{\partial [m_2]}{\partial\beta} 
 \end{equation} 
with $A>0$ a constant. Since both susceptibilities on the right-hand side are negative, a positive correlation between ceRNAs can emerge also in absence of miRNA-mediated cross-talk.

More recently, information theoretical quantities have been employed as a means to detect miRNA-mediated cross-talk. In \cite{sumazin}, for instance, the functional
\begin{equation}\label{di}
\Delta I([m_1],[\mu];[m_2])=\left\langle\left[I([m_1],[\mu])\right]\right\rangle_{[m_2]}-I([m_1],[\mu])
\end{equation} 
($I(x,y)$ denoting the mutual information of random variables $x$ and $y$, $\langle\cdots\rangle_z$ denoting the average with respect to the random variable $z$) has been proposed, with the rationale that if $\Delta I>0$ then the knowledge of $[m_2]$ increases the mutual dependence of $[m_1]$ and $[\mu]$, which can be interpreted as a signature of cross-talk between $m_1$ and $m_2$. 

In the previous section we have shown that neglecting molecular noise entirely and assuming that extrinsic transcriptional noise is the dominant source of stochasticity it is possible to characterize concentration fluctuations at stationarity using mean-field steady state equations once input noise is known.
We note, however, that under equations (\ref{ssse}) free ceRNA levels depend only on the levels of the miRNA they interact with, so that the joint probability distribution of the levels of the various molecular species involved can be factorized, e.g. for $N=2$ and $M=1$
\begin{equation}
P([m_1],[m_2]|[\mu])=P_1([m_1]|[\mu])\, P_2([m_2]|[\mu])~~.
\end{equation}
This in turn implies that, within this mean field steady state framework, the three-species correlation functions can also be factorized, i.e.
\begin{equation}
P([m_1],[m_2],[\mu])=P_1([m_1]|[\mu])\, P_2([m_2]|[\mu])P_{\mu}([\mu])~~,
\end{equation}
leading to $\Delta I=0$ independently of there being cross-talk or not.

Quite importantly, however, in a typical experimental output (e.g. by microarray or deep sequencing analysis) it is hard to disentangle the contribution of free and bound ceRNAs and the experimental readouts give proxies for the quantities
\begin{equation}\label{xp}
[m_i]_{xp}=[m_i]+[c_i]~~~~~,~~~~~
[\mu]_{xp}=[\mu] + [c_i]~~.
\end{equation}
Based on (\ref{sseq}), one has in particular
 \begin{gather}
 [m_i]_{xp}=c_i^\star \left[1 +\left(\frac{\sigma_i+\kappa_i}{d_i}-1\right)F_i([\mu]) \right]\\
 [\mu]_{xp}=[\mu]\left[ 1+\sum_i c_i^\star F_i([\mu])/\mu_{0,i} \right]
 \end{gather}
Note that if the lifetime of complexes ($\sigma_i+\kappa_i$) is shorter than that of ceRNAs $d_i$, as it is reasonable to expect, $[m_i]_{xp}$ is, like $[m_i]$, a decreasing function of $[\mu]$. One sees that, in general, it is not possible to express the experimental ceRNA levels in terms of the miRNA levels only. Therefore the quantity (\ref{di}) computed using the experimental readouts (\ref{xp}) can be different from zero. Again, however, this is not necessarily a signature of cross-talk. An argument is given in section {\it On the significance of conditional mutual information as a means to signal cross-talk} of the Supporting Text, where it is shown how non-zero values of $\Delta I([m_1]_{xp},[\mu]_{xp};[m_2]_{xp})$ can be obtained even in absence of stoichiometric complex degradation (and hence of cross-talk at stationarity). 

In summary, more refined detection methods are likely to be needed in order to identify cross-talk among ceRNAs from gene expression data.

\section*{Discussion}

Recent experimental studies have suggested that the miRNA-mediated competition between ceRNAs could constitute an additional level of post-transcriptional regulation, playing important roles in many biological contexts.   Trying to achieve a clear quantitative understanding of the emergence of this effect has been the goal of this work. We have presented a minimal, rate equation-based model that is able to describe the cross-talk arising from competition at steady state through a systematic analytical characterization of the sensitivity to small changes in the transcription rates. To keep mathematical complexities to a minimum, we have  adopted a coarse-grained view of the real biological process, even neglecting details of the miRNA-mediated regulation that could impact the emergence of cross-talk among ceRNAs. For instance, binding to the Argonaute/Ago protein (the catalytic component of the RISC) may represent a significant rate-limiting step \cite{risc_competition}, and  the competition for Ago has been shown to contribute to the emergence of ceRNA-ceRNA cross-talk \cite{sirna_competition}.

The emerging scenario, valid in the linear response regime, is rather rich and complex. Interestingly, the competitive interactions can give rise to a rather selective communication channel: only ceRNAs in an intermediate, susceptible regime are responsive to miRNA perturbations and significantly contribute to diluting the strength of the interaction. Thus, even in case of a dense miRNA-ceRNA network, the resulting ceRNA-ceRNA cross-talk pattern may be rather sparse. Moreover, interactions switch on only in specific ranges of miRNA concentrations, so that the structure of the emergent ceRNA-ceRNA network can adjust in response to variations in the miRNA levels. Perhaps unexpectedly, heterogeneity of kinetic parameters can give rise both to symmetric and asymmetric couplings. Furthermore, an analogous cross-talk scenario emerges between different miRNA species sharing the same target RNA. And, finally, the topology of the ceRNA-miRNA network may play an important role as strong correlations in connectivity in that network can enhance the ceRNA-ceRNA cross-talk. 

The above picture requires that miRNA-ceRNA complexes decay, at least partially, through a stoichiometric channel of degradation: for purely catalytic decay no cross-talk is possible at stationarity, and perturbations of transcription rates only cause a transient response. Dynamical effects may nevertheless play an important role on the time-scales of many cellular processes, and will be explored in a forthcoming work.

In order to evaluate the robustness of the miRNA-mediated coupling, we have also performed a basic analysis of the impact of noise. Assuming extrinsic transcriptional noise as the dominant source of stochasticity, we estimated level fluctuations in the ceRNA-miRNA networks at steady state, again obtaining expressions valid in the linear response regime. It turns out that miRNA-mediated cross-talk, coupled with correlated transcriptional inputs, represents a powerful noise processing mechanism that can lead to either noise reduction or amplification. It is interesting to observe that a circuit displaying specific transcriptional correlations has been discussed in  \cite{cesana}, where a muscle-specific miRNA (miR-133b) embedded in a non-coding transcript (linc-MD1) has been identified. Clearly, linc-MD1's transcript acts as a very efficient decoy for miR-133b. A theory for this case is worked out in section {\it The miRNA-decoy transcript} of the Supporting Text. 

It would be important to carry the analysis of the role of noise beyond the steps discussed here. Post-transcriptional regulation based on stoichiometric repression has been shown to cause large intrinsic fluctuations in intermediate regimes of repression \cite{mehta}, effectively posing a limit to the possibility of having an efficient quantitative signaling between ceRNAs. Our analysis suggests on the other hand that cross-talk mediated by large number of miRNAs might be more robust. A more thorough mathematical/computational analysis, including molecular noise, may be able to shed light on this important aspect.

We have finally shown that non-trivial correlations among ceRNAs can emerge in experimental readouts due to transcriptional fluctuations even in absence of miRNA-mediated cross-talk. 

\subsection*{Acknowledgments}

While completing this manuscript we learned that C. Bosia, A. Pagnani and R. Zecchina  have independently studied the same problem,  reporting results which are consistent with those obtained here. We thank C. Bosia, I. Bozzoni, M. Caselle, A. Martirosyan, P. Mehta, A. Pagnani, R. Zecchina for stimulating discussions.

\newpage

\section*{Supporting Text}

\subsection*{Derivation of miRNA steady-state concentration}

An approximate, explicit expression for the steady state miRNA level  can be derived starting from (9) of the Main Text. For simplicity, let us consider the case of $N=2$ ceRNAs, in which (8) of the Main Text reduces to the equation
\begin{equation}
\label{mu_eq1}
[\mu] \left[ \delta + b_1 z_1 F_1([\mu])+ b_2 z_2 F_2([\mu])\right]=\beta~~.
\end{equation}
We can work out its solutions explicitly depending on the regimes to which the ceRNAs belong by inserting (9) of the Main Text into (\ref{mu_eq1}), using the relation $w_i=z_i\, \mu_{0,i}$ and keeping only linear terms in $[\mu]$. One finds the following results: 
\begin{description}
\item[(a) $m_1,m_2\in \mathcal{F}$] : 
\begin{equation}
[\mu]\simeq\frac{\beta}{\delta+\sum_i b_i  z_i}
\end{equation}
\item[(b) $m_1,m_2\in \mathcal{S}$] : 
\begin{equation}
[\mu]\simeq\frac{\beta-\frac{1}{4}\sum_i b_iw_i}{\delta+\frac{1}{4}\sum_i b_i  z_i}
\end{equation}
\item[(c) $m_1,m_2\in \mathcal{B}$] : 
\begin{equation}
[\mu]\simeq\frac{\beta-\sum_i b_iw_i}{\delta}
\end{equation}
\item[(d) $m_1\in \mathcal{B},\,\,\,m_2\in \mathcal{S}$] : 
\begin{equation}
[\mu]\simeq\frac{\beta-b_1w_1-\frac{1}{4}b_2w_2}{\delta+\frac{1}{4}b_2  z_2}
\end{equation}
\item[(e) $m_1\in \mathcal{F},\,\,\,m_2\in \mathcal{B}$] : 
\begin{equation}
[\mu]\simeq\frac{\beta-b_2w_2}{\delta+b_1  z_1}
\end{equation}
\item[(f) $m_1\in \mathcal{F},\,\,\,m_2\in \mathcal{S}$] : 
\begin{equation}
[\mu]\simeq\frac{\beta-\frac{1}{4}b_2w_2}{\delta+b_1  z_1+\frac{1}{4}b_2  z_2}
\end{equation}
\end{description}
Extending to the general case of $N$ ceRNAs we conclude that
\begin{equation}
[\mu]\simeq\frac{\beta-\sum_{i\in\mathcal{B}}b_i w_i-\frac{1}{4}\sum_{i\in \mathcal{S}} b_i w_i}{\delta+\sum_{i\in\mathcal{F}} b_i z_i+\frac{1}{4}\sum_{i\in \mathcal{S}} b_i z_i}~~,
\end{equation}

\subsection*{The mirror system: one target, $M$ miRNA species}

The dual system in which $M$ miRNA species, labeled $\mu_\alpha$ ($\alpha=1,\ldots,M$), target the same RNA $m$ (to avoid confusion we will keep referring to it as a ceRNA even though in this case it is not really competing, being the only target species), can be worked out in full analogy with the case discussed above. In particular, defining $\mu^\star_{\alpha}=\beta_\alpha/\delta_\alpha$, we have that, at stationarity, the level of free miRNA species is given by
\begin{equation}
[\mu_\alpha]=\mu^\star_\alpha F_\alpha([m])~~,
\end{equation}
where
\begin{equation}
F_\alpha=\frac{m_{0,\alpha}}{[m]+m_{0,\alpha}}~~~,~~~
m_{0,\alpha}=\frac{\delta_\alpha}{k_\alpha^+}(1+\psi_\alpha)
\end{equation}
with $\psi_\alpha=(k_\alpha^-+\kappa_\alpha)/\sigma_\alpha$. (Note that now rates carry the index of the corresponding miRNA involved.) The free ceRNA level on the other hand results from the algebraic equation
\begin{equation}
\label{eq_m}
[m] \left[ d+\sum_\alpha \beta_\alpha \zeta_\alpha F_\alpha([m])\right]=b ~~,
\end{equation}
where $\zeta_\alpha=(\sigma_\alpha+\kappa_\alpha)/(m_{0,\alpha}\sigma_\alpha)$. As before, each $m_{0,\alpha}$ can be interpreted as reference level for the target which can be used to separate different regimes for the miRNA species. Borrowing the terminology used in the previous case, we have
\begin{equation}
\label{expansion_mir}
F_\alpha([m])\simeq 
\begin{cases}
1-[m]/m_{0\alpha}& \alpha\in \mathcal{F} \\
\frac{1}{2}-([m]-m_{0,\alpha})/(4m_{0,\alpha})& \alpha\in \mathcal{S}\\
m_{0,\alpha}/[m]& \alpha\in\mathcal{B}
\end{cases}
\end{equation}
where $[m] \ll m_{0,\alpha}$ for a `free' miRNA, $[m] \simeq m_{0,\alpha}$ for a `susceptible' miRNA, and $[m] \gg m_{0,\alpha}$ for a `bound miRNA. In turn, the level of free ceRNA is given by
\begin{equation}
[m]\simeq\frac{b-\sum_{\alpha\in\mathcal{B}}\beta_\alpha \omega_\alpha-\frac{1}{4}\sum_{\alpha\in \mathcal{S}}
\beta_\alpha \omega_\alpha}{d+\sum_{\alpha\in \mathcal{F}} \beta_\alpha \zeta_\alpha+\frac{1}{4}\sum_{\alpha\in \mathcal{S}} \beta_\alpha \zeta_\alpha}
\end{equation}
where $\omega_\alpha=\zeta_\alpha\, m_{0,\alpha}=(\sigma_\alpha+\kappa_\alpha)/\sigma_\alpha$, while for the susceptibility we obtain
\begin{equation}
\chi_{\alpha \gamma}\equiv\frac{\partial [\mu_\alpha]}{\partial \beta_\gamma}=\frac{1}{\delta_\alpha}\left[F_\alpha([m])\delta_{\alpha\gamma}+\frac{\beta_\alpha  \zeta_\gamma \chi_{m,m}}{4[m]}W_{R(\alpha),R(\gamma)}\right]
\end{equation}
where $\chi_{m,m}=\frac{\partial [m]}{\partial b}$ and the matrix $\widehat{W}$ is the same as in (14) of the Main Text with $[m]$, $m_{0,\alpha}$ and $m_{0,\gamma}$ replacing respectively $[\mu]$, $\mu_{0,i}$ and $\mu_{0,j}$.

\begin{figure}
\begin{center}
\includegraphics[width=3.25in]{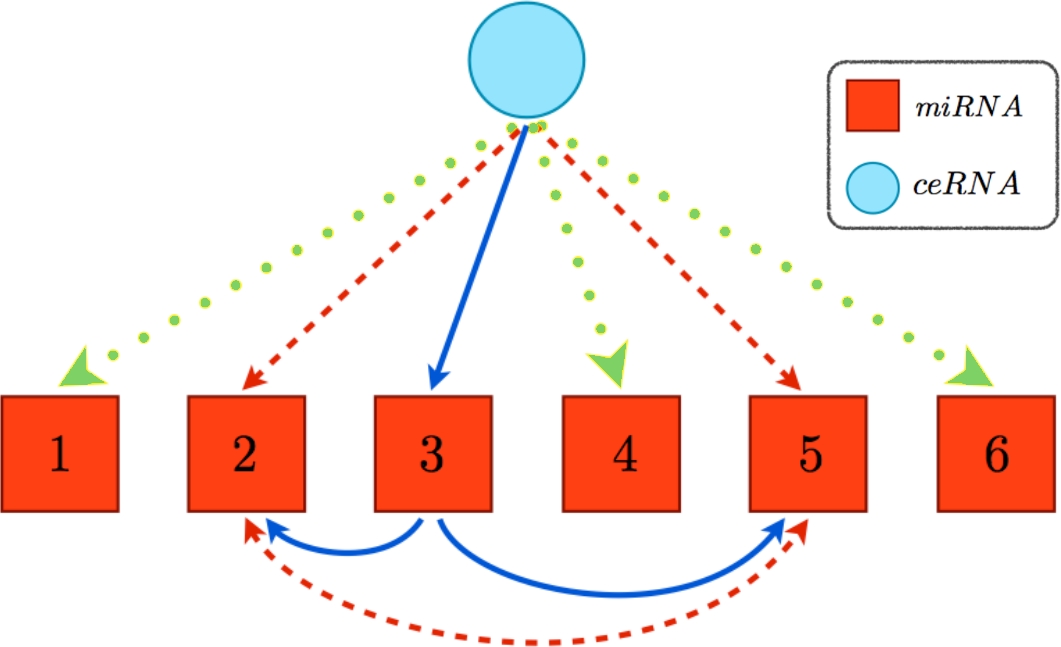}
\caption{\label{manymir} Schematic representation of a system of one target RNA and $M=6$ miRNAs species. In this case miRNA 3 is Bound, miRNAs 2 and 5 are Susceptible and the remaining are Free from the target RNA. Cross-talk interactions pattern is derived analogously to the dual case discussed in the main text: symmetrical cross-talk interactions emerge between miRNA 2 and 5 and asymmetrical interactions emerge from miRNA 3 to miRNAs 2 and 5.}
\end{center}
\end{figure}

Therefore the cross-talk that is established between miRNAs is, as before, selectively turned on only for species lying in particular regimes, defined by the free ceRNA level. In complete analogy to the dual system analyized in the main text, two types of effective interactions arise: the first one is symmetric encodes the response of a miRNA in the $\mathcal{S}$-regime to a perturbation of another miRNA in the $\mathcal{S}$-regime; the second one is asymmetric and encodes the response of a miRNA in the  $\mathcal{S}$-regime to a perturbation of a miRNA in the $\mathcal{B}$-regime. An example of a pattern of interactions between miRNAs is shown in Figure \ref{manymir}. 
Note that the intensity of the cross-talk is modulated by the factor $\zeta_\gamma$ and increases when the rate of catalytic degradation increases. If $\sigma_\alpha \rightarrow 0$ for all $\alpha$ implies $m_{0,\alpha}\to\infty$: in this case,  all miRNA species lie in the $\mathcal{F}$-regime and no cross-talk is possible at steady state.

\subsection*{The role of topology}

Network topology can play an important role as a cross-talk enhancer. In specific, we will now argue that ceRNA-ceRNA interactions can be mediated by a large number of miRNA species which individually would only weakly dampen ceRNA levels. We consider a diluted network described by an adjacency matrix $\{A_{i\alpha}\}$ such that
\begin{equation}
A_{i\alpha}=
\begin{cases}
1\quad  \text{if ceRNA $m_i$ is targeted by miRNA $\mu_\alpha$}\\
0 \quad  \text{otherwise}\\
\end{cases}
\end{equation}
making the following simplifying assumptions: (a) the network is kinetically homogeneous, i.e. rates are the same for all ceRNAs, so that $\mu_{0,i\alpha}=\mu_0$ for each $i$ and $\alpha$ and $b_i=b$, $d_i=d$ for each $i$; (b) miRNA levels are uniform, i.e. $[\mu_\alpha]=[\mu]$ for all $\alpha$; (c)  $[\mu]/\mu_0\equiv t\ll 1$, so that all ceRNA species are in the $\mathcal{F}$-regime with respect to any miRNA ($i\in\mathcal{F}(\alpha)\,\,\forall i,\alpha$).

Consider a pair of ceRNAs $m_i$ and $m_j$, targeted respectively by $n_i=\sum_\alpha A_{i\alpha}$ and $n_j=\sum_\alpha A_{j\alpha}$ miRNA species, $n_{ij}=\sum_\alpha A_{i\alpha}A_{j\alpha}$ of which are in common. In this case, the ceRNA concentration reads $m_i=m^\star/(1+n_i t)$ (with $m^\star=b/d$) and the cross-susceptibility (28) of the Main Text turns out to be given by
\begin{equation}
 \chi_{ij,\alpha}=
\begin{cases} 
 \frac{1}{d K_\alpha}\,\frac{t}{[1+t(n_j-1)][1+t(n_i-1)]^2}\quad \text{if}&\quad A_{i\alpha}A_{j\alpha}=1\\
 0 &\quad \textit{otherwise}
\end{cases}
\end{equation}
where
$K_\alpha\simeq [\delta/(z b)+\sum_{k\in \alpha} (1+t n_k)^{-1}]$ and $z$ is defined by the fact that $z_{i\alpha}=\frac{\sigma}{(\sigma+\kappa)\mu_{0}}=z$ for each $i$ and $\alpha$. (The notation $k\in \alpha$ indicates all ceRNAs interacting with miRNA $\mu_\alpha$.) As expected, the dilution increases upon increasing the number of ceRNAs interacting with a given miRNA species $\mu_\alpha$ (each of them add a positive term $(1+t n_k)^{-1}$ to $K_\alpha$ thus making it larger) and upon increasing $n_i$ and $n_j$, since
\begin{equation}
\chi^\alpha_{ij}\propto \frac{1}{n_j n_i^2}~~~~~\,\,(n_i,n_j \gg 1/t)~~.
\end{equation}

\begin{figure}
\begin{center}
\includegraphics[width=3.25in]{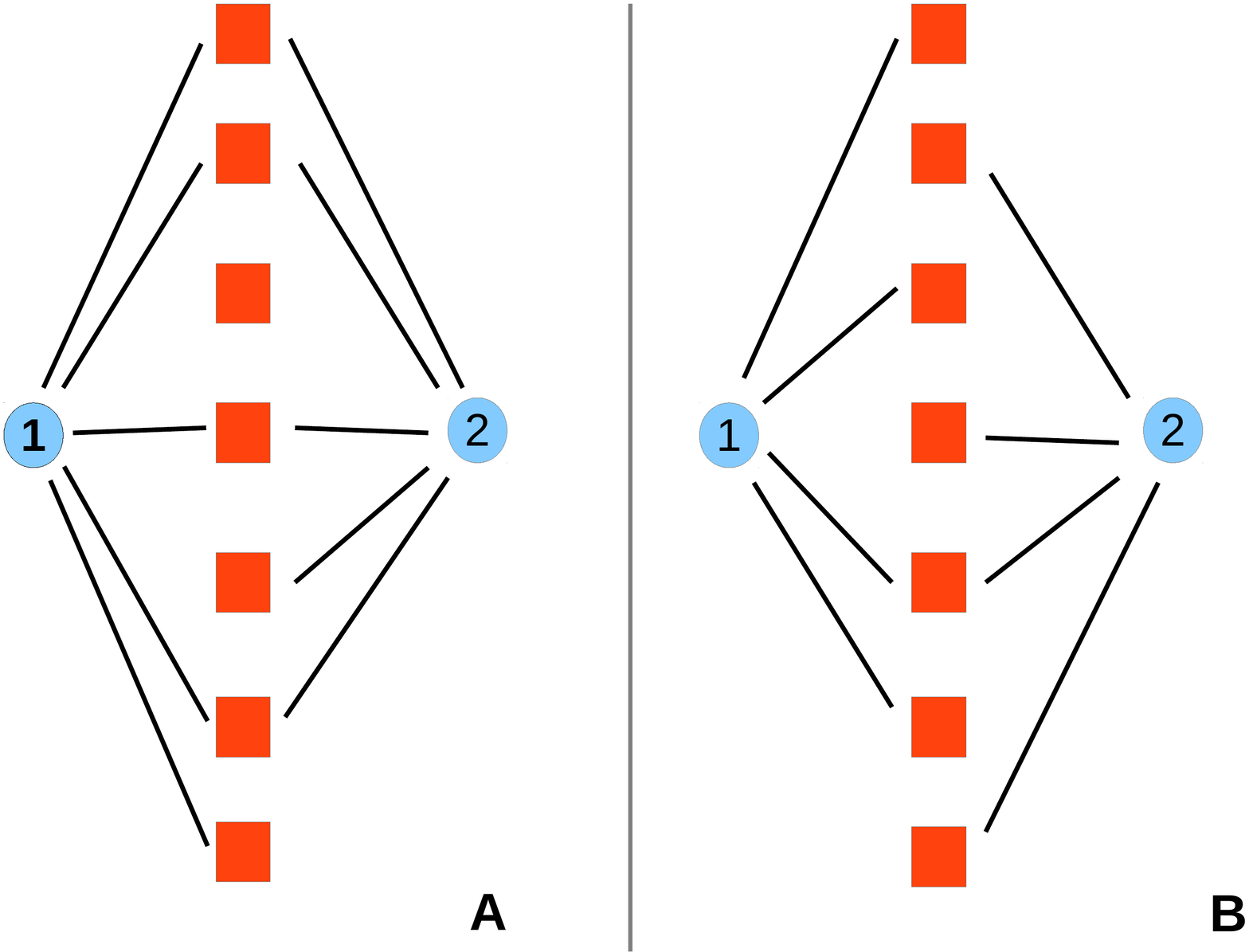}
\caption{\label{topology} Two examples of different network structures with $N=2$ ceRNAs (blue circles) and $M=7$ miRNAs (red squares). A) A highly correlated network structure where ceRNAs share almost all of their regulators ($n_1=n_2=5,\,\,n_{12}=4$). B) A poorly correlated structure where ceRNAs share a small fraction of their regulators ($n_1=n_2=4,\,\,n_{12}=1$). Cross-talk will tipically be much stronger in A than in B.
}
\end{center}
\end{figure}

Consider now the particular case of a regular bipartite network with fixed ceRNA and miRNA connectivity so that $n_i=n$ for each $i$ and $\nu_\alpha\equiv\sum_i A_{i\alpha}=\nu$ for each $\alpha$. Setting 
\begin{equation}
K_\alpha=K= \frac{\delta}{z b}+\frac{\nu}{1+t n}
\end{equation}
for all $\alpha$ we clearly see that now each miRNA species contributes equally  to the overall susceptibility, i.e, $\chi_{ij,\alpha}=\chi_0$ for all $i$ and $j$ targeted by $\mu_\alpha$ with
\begin{equation}
 \chi_0=\frac{1}{d K}\,\frac{t}{[1+t(n-1)][1+t(n-1)]^2}~~,
\end{equation}
while the overall susceptibility is given by $\chi_{ij}= n_{ij}\chi_0$. The contribution of a single miRNA to the overall susceptibilities will depend on the value of $t$. In particular, one easily sees that
\begin{equation}
\chi_0=
\begin{cases}
\frac{t}{d K}\sim \mathcal{O}(\frac{\epsilon}{n}) \quad \text{for $t\ll 1/n$}\\
\frac{1}{d K n}\sim \mathcal{O}(\frac{1}{n})\quad \text{for $t\simeq 1/n$}\\
\frac{1}{d K t^2 n^3}\sim\mathcal{O}(\frac{\epsilon}{n}) \quad \text{for  $t\gg 1/n$}\\
\end{cases}
\end{equation}
Generalizing the Free, Susceptible and Bound regimes, one realizes that the case $t\ll 1/n$ (resp. $t\simeq 1/n$ and $t\gg 1/n$) describes a ceRNA that is `globally free' (resp. `globally susceptible' and `globally bound') with respect to the overall miRNA population. We therefore conclude that $\chi_{ij}$
\begin{enumerate}
\item[(i)] increases with the number $n_{ij}$ of miRNA species shared by the ceRNAs $m_i$ and $m_j$;
\item[(ii)] decreases if the shared miRNAs have many other targets;
\item[(iii)] peaks when ceRNAs are `globally susceptible' to the overall miRNA population, and it can be of the same order of magnitude as the self-susceptibility, i.e. $\mathcal{O}(1/d)$, when  $n_{ij}\simeq n$.
\end{enumerate}
Perhaps most remarkably, the cross-talk can be effective even among ceRNAs that are in the Free regime with respect to individual miRNAs, provided they are commonly targeted by a large number of miRNA species thus becoming 'globally susceptible'.  However, in order to achieve efficient cross-talk strong correlations in the network connectivity are needed (large $n_{ij}$): highly clustered networks can allow for much stronger cross-talk than random graphs (see Figure \ref{topology}).

\subsection*{The miRNA-decoy transcript}

Many miRNAs (possibly about $50\%$ of the total \cite{bartel}) are hosted in non-coding genes whose transcript can incur a dual fate: after transcription, the precursors can either be processed into mature miRNAs through a series of steps involving proteins DROSHA and DICER, or they can reach the cytoplasm unprocessed in the form of long non-coding RNAs (lncRNAs). The RNA sequence close to the sites corresponding to the miRNA presents a region with a sequence that is almost complementary to that of miRNA. These proximal strings allow for the miRNA precursor (pri-miRNA) to take on the peculiar hairpin structure that is essential  for the recognition by the processing proteins and thus for miRNA maturation \cite{bartel2}. It also follows, however, that the RNA sequence close to the miRNA necessarily contains a good potential binding site for the miRNA itself. When matured into lncRNAs, such transcripts are thus targeted by the miRNA and represent efficient `miRNA traps' or decoys, through which the population of miRNAs available for target repression can be regulated. The above miRNA-decoy mechanism can be modeled with following processes (see also Fig. \ref{spongemir}): 
\begin{figure}
\begin{center}
\includegraphics[width=2.25in]{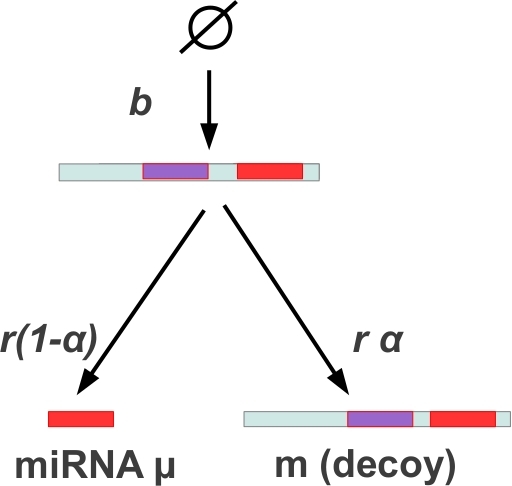}
\caption{\label{spongemir}  Schematic representation of the model of a miRNA-decoy transcript.}
\end{center}
\end{figure}
\begin{equation}
\emptyset \overset{b}\rightarrow q ~~~~~~~~~~~~~
q \overset{r \alpha}\rightarrow m ~~~~~~~~~~~~~
q \overset{r (1-\alpha)}\rightarrow \mu ~~,
\end{equation}
including transcription of the long non-coding RNA $q$ at rate $b$, transport of $q$ to the cytoplasm with processing into mature miRNA $\mu$ at rate $(1-\alpha) r$, and transport of $q$ to the cytoplasm $\alpha r$. The quantity $1-\alpha\in[0,1]$ thus gives the fraction of miRNA produced over the total number of transcribed RNAs.

\begin{figure}
\begin{center}
\includegraphics[width=3.25in]{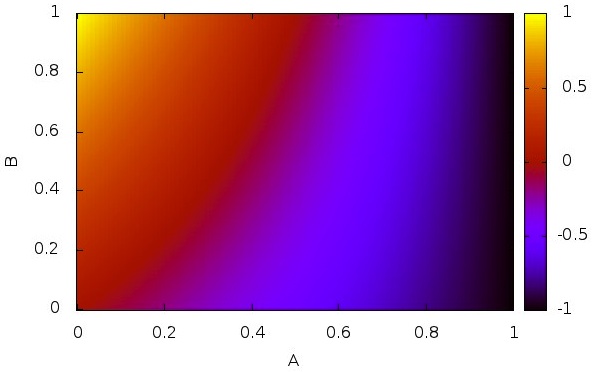}
\caption{\label{pearson_extr} Pearson correlation coefficient between the production rate of miRNA $\mu$ and of decoy $m$, for different values of the processing noise level ($A\equiv \sigma^2_\alpha/[\alpha(1-\alpha)]$ on the \textit{x} axis) and of the transcription noise level ($B\equiv \sigma^2_b/b^2$ on the \textit{y} axis). High level of processing noise gives rise to negative correlations, while low level of processing noise and high level of transcriptional noise result in positive correlations.
}
\end{center}
\end{figure}

At stationarity, the miRNA and the lncRNA $m$ are produced at constant rates according to
\begin{gather}
\dot{m}=b\alpha  \\
\dot{\mu}=b(1-\alpha)
\end{gather}
If noise affects both the transcription rate $b$ and the processing efficiency $\alpha$ (taking again Gaussian distributions with means $\overline{b}$ and $\overline{\alpha}$ and variances $\sigma^2_b$ and $\sigma^2_\alpha$, respectively), the covariance between production rates is easily seen to be given by
\begin{equation}
\overline{\dot{m} \, \dot{\mu}}-\overline{\dot{m}} \,\,\overline{\dot{\mu}}=\sigma^2_b(\overline{\alpha}-\overline{\alpha^2})-\sigma^2_\alpha \overline{b^2}
\end{equation}
Hence noise in $b$ and $\alpha$ induces noise at the level of molecular concentrations, yielding either positive or negative correlations between the steady state production rates of the miRNA $\mu$ and of the decoy $m$  as shown in Figure \ref{pearson_extr}. 
(Clearly, this conclusion holds as long as the noise on $\alpha$ is sufficiently small, or $A$ is not too close to 1.)

These correlations, in turn, can result in a change of steady state fluctuations of other competing RNAs through the usual miRNA-mediated channels. In the case of muscle differentiation discussed in \cite{cesana}, large levels of noise at the transcriptional or at the processing level could be exploited in order to increase cell variability and give rise to the differentiation program. Such a mechanism could be shared by other miRNA genes representing a widespread network motif.

\subsection*{On the significance of the conditional mutual information as a means to signal cross-talk}

Consider a system $(t,m,\mu)$ of $2$ ceRNAs (a target $t$ and a modulator $m$ and $N$ background targets) and one miRNA $\mu$, subject to transcriptional fluctuations. Let us say that the experimental readouts concern the quantities
\begin{gather}
[m]_{xp}=[m]+[c_m]\\
[t]_{xp}=[t]+[c_t]\\
[\mu]_{xp}=[\mu]+[c_m]+[c_t]
\end{gather}
where $[c_t]$ and $[c_m]$ represent the levels of miRNA-target and miRNA-modulator complexes, respectively. Suppose that both complexes decay catalytically, i.e. that the rates of stoichiometric complex degradation $\sigma_m=\sigma_t=0$. In such conditions no cross-talk is possible at steady state. Furthermore, let us assume that the transcription rates $b_t$, $b_m$, and $\beta$ are drawn from a probability distribution $P_0(b_t,b_m,\beta)$ such that
\begin{equation}
P_0(b_t,b_m,\beta)\equiv P(b_t,\beta)\delta(b_m-k)
\end{equation}
with $P$ an unspecified probability distribution with finite covariance (i.e., that the target and miRNA transcription rates are random variables while the modulator transcription rate is fixed at $k$). We want to show that, in this case, $\Delta I([t]_{xp},[\mu]_{xp};[m]_{xp})>0$  (with $\Delta I$ defined in (37) of the Main Text) necessarily. This would imply that the condition $\Delta I([t]_{xp},[\mu]_{xp};[m]_{xp})>0$ cannot be considered as a sufficient condition for cross-talk, since knowledge of $[m]_{xp}$ can increase the mutual dependence between $[\mu]_{xp}$ and $[t]_{xp}$ even in absence of cross-talk. 

To see this, note that the measured steady state levels are stochastic variables which depend on the transcription rates as
\begin{gather}
[m]_{xp}=f_m(\beta)  \label{emme} \\
[\mu]_{xp}=f_\mu(b_t,\beta) \label{muuuu} \\
[t]_{xp}=f_t(b_t,\beta) \label{tiiii}
\end{gather}
(with $f_m$, $f_\mu$ and $f_t$ unspecified functions). Now let us focus on (\ref{emme}) and (\ref{muuuu}). Given their monotonicity with respect to each of the variables on which they depend, they can be inverted:
\begin{gather}
\beta=f^{-1}_m([m]_{xp})\\
b_t=f^{-1}_\mu([\mu]_{xp},\beta)
\end{gather}
Hence it is possible to express $[t]_{xp}$ as a function of $[m]_{xp}$ and $[\mu]_{xp}$ directly: $[t]_{xp} =h([m]_{xp},[\mu]_{xp})$. In other terms, one finds a deterministic dependence of $[t]_{xp}$ on $[m]_{xp}$. This implies that for each fixed $[m]_{xp}$ the mutual information between $[t]_{xp}$ and $[\mu]_{xp}$ diverges. As a consequence, their mutual information averaged over $[m]_{xp}$, $\avg{I ([t]_{xp},[\mu]_{xp})}_{[m]_{xp}}$, diverges as well. At the same time, however, the mutual information between $[t]_{xp}$ and $[\mu]_{xp}$ stays finite due to the noise on $b_t$ and $\beta$. Hence
\begin{multline}
\Delta I([t]_{xp},[\mu]_{xp};[m]_{xp})\equiv \\ \equiv \avg{I ([t]_{xp},[\mu]_{xp})}_{[m]_{xp}}- I([t]_{xp},[\mu]_{xp})>0~~
\end{multline}
necessarily.

\end{document}